\documentclass[fleqn,usenatbib]{mnras}

\usepackage{newtxtext,newtxmath}

\usepackage[T1]{fontenc}

\DeclareRobustCommand{\VAN}[3]{#2}
\let\VANthebibliography\thebibliography
\def\thebibliography{\DeclareRobustCommand{\VAN}[3]{##3}\VANthebibliography}

\usepackage{graphicx}	
\usepackage{amsmath}	
\usepackage{amssymb}	
\usepackage{multirow}   
\usepackage{tikz,xcolor,hyperref} 

\definecolor{lime}{HTML}{A6CE39}
\DeclareRobustCommand{\orcidicon}{%
	\begin{tikzpicture}
	\draw[lime, fill=lime] (0,0) 
	circle [radius=0.16] 
	node[white] {{\fontfamily{qag}\selectfont \tiny ID}};
	\draw[white, fill=white] (-0.0625,0.095) 
	circle [radius=0.007];
	\end{tikzpicture}
	\hspace{-2mm}
}

\foreach \x in {A, ..., Z}{%
	\expandafter\xdef\csname orcid\x\endcsname{\noexpand\href{https://orcid.org/\csname orcidauthor\x\endcsname}{\noexpand\orcidicon}}
}

\title[XCS follow-up of eFEDS selected cluster candidates]{The {\em XMM} Cluster Survey: An independent demonstration of the fidelity of the eFEDS galaxy cluster data products and implications for future studies}

\author[D. J. Turner et al.]
{D. J. Turner$^{1}$\thanks{E-mail: david.turner@sussex.ac.uk (DJT)}\orcidA{},
P. A. Giles$^{1}$\orcidB{},
A. K. Romer$^{1}$\orcidC{},
R. Wilkinson$^{1}$\orcidD{},
E. W. Upsdell$^{1}$\orcidG{},
M. Klein$^{2}$,
\newauthor
P. T. P. Viana$^{3,4}$\orcidE{},
M. Hilton$^{5,6}$\orcidJ{},
S. Bhargava$^{7}$,
C. A. Collins$^{8}$,
R. G. Mann$^{9}$\orcidH{},
M. Sahl\'en$^{19}$\orcidI{},
J. P. Stott$^{11}$\orcidF{}
\\
$^{1}$Department of Physics and Astronomy, University of Sussex, Brighton, BN1 9QH, UK\\
$^{2}$Faculty of Physics, Ludwig-Maximilians-Universität, Scheinerstr. 1, 81679, Munich, Germany\\
$^{3}$Instituto de Astrof\'isica e Ci\^{e}ncias do Espa\c co, Universidade do Porto, CAUP, Rua das Estrelas, 4150-762 Porto, Portugal \\
$^{4}$Departamento de F\'isica e Astronomia, Faculdade de Ci\^{e}ncias, Universidade do Porto, Rua do Campo Alegre, 687, 4169-007 Porto, Portugal \\
$^{5}$Astrophysics Research Centre, University of KwaZulu-Natal, Westville Campus, Durban 4041, SA \\
$^{6}$School of Mathematics, Statistics, and Computer Science, University of KwaZulu-Natal, Westville Campus, Durban 4041, SA \\
$^{7}$Département d’Astrophysique, CEA Paris-Saclay, 91190 Gif-sur-Yvette, France\\
$^{8}$Astrophysics Research Institute, Liverpool John Moores University, Liverpool Science Park, 146 Brownlow Hill, Liverpool L3 5RF, UK \\
$^{9}$Institute for Astronomy, University of Edinburgh, Royal Observatory, Blackford Hill, Edinburgh EH9 3HJ, UK \\
$^{10}$Theoretical Astrophysics, Department of Physics and Astronomy, Uppsala University, Box 516, SE- 751 20 Uppsala, Sweden\\
$^{11}$Department of Physics, Lancaster University, Lancaster LA1 4YB, UK 
}

\date{Accepted XXX. Received YYY; in original form ZZZ}

\pubyear{2021}

\begin{document}
\label{firstpage}
\pagerange{\pageref{firstpage}--\pageref{lastpage}}
\maketitle

\begin{abstract}
We present the first comparison between properties of clusters of galaxies detected by the eROSITA Final Equatorial-Depth Survey (eFEDS) and the {\em XMM} Cluster Survey (XCS). We have compared, in an ensemble fashion, properties from the eFEDS X-ray cluster catalogue with those from the Ultimate {\bf X}MM e{\bf X}traga{\bf L}actic (XXL) survey project (XXL-100-GC). We find the distributions of redshift and X-ray temperature ($T_{\rm X}$) to be broadly similar between the two surveys, with a larger proportion of clusters above 4\,keV in the XXL-100-GC sample. We find 62 eFEDS cluster candidates with {\em XMM} data (eFEDS-{\em XMM} sample); 10 do not have good enough {\em XMM} data to confirm or deny, 11 are classed as sample contaminants, and 4 have their X-ray flux contaminated by another source. The majority of eFEDS-{\em XMM} sources have a longer exposure in {\em XMM} than eFEDS, and the majority of eFEDS positions are within 100~kpc of XCS positions. Our eFEDS-XCS sample of 37 clusters is used to calculate minimum sample contamination fractions of ${\sim}$18\% and ${\sim}$9\% in the eFEDS X-ray and optically confirmed samples respectively, in general agreement with eFEDS findings. We compare 29 X-ray luminosities ($L_{\rm X}$) measured by eFEDS and XCS, which are in excellent agreement. Eight clusters have a $T_{\rm X}$ measured by {\em XMM} and {\em eROSITA}, and we find that {\em XMM} temperatures are 25$\pm$9\% larger than their {\em eROSITA} counterparts. Finally, we construct $L_{\rm X}$ - $T_{\rm X}$ scaling relations based on eFEDS and XCS measurements, which are in tension; the tension is decreased when we measure a third scaling relation with calibrated XCS temperatures.

\end{abstract}




\section{Introduction}
X-ray observations of clusters of galaxies provide insights into various aspects of astrophysics  \cite[e.g.,][]{,hitomi16turb,bhargava20,sanders21} and cosmology \cite[e.g.,][]{, vikhlinincosmo, otherxraycosmo}.
Clusters are among the largest gravitationally bound structures in the Universe and consist of a dark matter halo, the intra-cluster medium (ICM), and the component galaxies. The ICM is a high-temperature, low-density plasma that emits strongly in the X-ray band, with both continuum and emission line components.

The {\em eROSITA} instrument mounted on the joint Russian-German Spectrum-Roentgen-Gamma \citep[SRG,][]{missionpaper} mission will contribute significantly to X-ray cluster astrophysics and cosmology. Its large field of view (${\sim}$1~deg), sensitivity, and energy resolution combine to make it a revolutionary new instrument. The final {\em eROSITA} All-Sky Survey (eRASS) is predicted to detect approximately 100,000 galaxy clusters above a mass of $5\times10^{13}$h$^{-1}$M$_{\odot}$ \citep[][]{erass_numclusters}. The data sharing agreement between the German and Russian consortiums that funded {\em eROSITA} involves sharing the sky equally. Most of these clusters will be accompanied by an X-ray luminosity ($L_{\rm X}$) measurement, and roughly 20\% \citep[][]{efedsclustercat} of the observations will yield an X-ray temperature ($T_{\rm X}$) measurement. However, apart from a handful of the highest flux clusters, it will not be possible to measure masses via the hydrostatic technique directly from eRASS data. Therefore, until the all sky survey is complete, it will be necessary to supplement the eRASS cluster catalogue with mass measurements from the current generation of X-ray telescopes (i.e. {\em XMM}, {\em Chandra}) in order to maximise the scientific yield. After the all sky survey is complete, {\em eROSITA} pointed observations of clusters, will produce some hydrostatic mass estimates, as demonstrated \cite{pointysanders}.

The aim of this paper is to explore potential synergies between eRASS cluster catalogues and the data in the {\em XMM}-Newton public archive, and to probe calibration considerations required for such analyses. The {\em eROSITA} and {\em XMM} telescopes have different characteristics that allow them to complement one another, some of which (such as the effective area at different energies, and the background level) were explored by \cite{missionpaper}. Comparisons between the on-axis effective areas of the combined {\em XMM} cameras (PN, MOS1, and MOS2) and the combined {\em eROSITA} telescope modules show that the effective areas are effectively equal between ${\sim}$0.5-2.0~keV, though outside this range {\em XMM} has an advantage. Here {\em XMM} complements {\em eROSITA} in that it will observe more source emission at higher energies, which could improve constraints on spectroscopic X-ray measurements of temperature and luminosity. The larger field of view of {\em eROSITA} ensures that its grasp (the product of effective area and observing area) is significantly greater than {\em XMM}'s below ${\sim}3.5$~keV, though above that energy {\em XMM}'s grasp is greater. Comparisons between the background levels of a subset of {\em eROSITA}'s telescope modules and the {\em XMM-Newton} cameras using a simultaneous observation of NLS1 1H0707-495 \citep[][]{simulback} revealed that, although the {\em eROSITA} background is higher than pre-launch predictions, it is generally lower than {\em XMM} and more temporally stable. Soft-proton flaring does not significantly impact {\em eROSITA}, giving it an advantage over {\em XMM} in this regard, and possibly allowing it to locate more low-flux sources (such as low surface-brightness galaxy clusters).

In this work we will make use of the recent release of the {\em eROSITA} Final Equatorial-Depth Survey \citep[eFEDS,][]{efedscat}. The eFEDS field covers approximately 140 square degrees of the equatorial ($-2.5^{\circ}<\delta<6.0^{\circ}$) sky. It intersects with several optical/near-IR photometric and/or spectroscopic surveys, including the Hyper Suprime-Cam Subaru Strategic Program \citep[HSC SSP,][]{hscsurvey}, the Galaxy and Mass Assembly survey \citep[GAMA,][]{gamasurvey}, and the Sloan Digital Sky Survey \citep[SDSS,][]{sdss}. We make an indirect comparison to a similar X-ray survey using the XXL-100-GC catalogue \citep[][]{xxlgc100}, then make direct comparisons to XCS measurements.

We wish to ascertain the level of sample contamination in the eFEDS cluster catalogue, compare the central coordinates of the detected clusters to those measured by the {\em XMM} Cluster Survey \citep[XCS, ][]{xcsfoundation}, and verify the accuracy of $L_{\rm{X}}$ and $T_{\rm{X}}$ measurements. 
As eFEDS is the same depth as the final eRASS, the accuracy of these measurements have implications for cosmological studies based on eRASS cluster detection (using weak lensing masses and X-ray luminosities for the mass observable relation). They will also impact studies based on optical or near-infrared detection (as luminosities can be used to explore scatter in the mass observable relations), and astrophysical studies of cluster luminosity-temperature relations to study the evolution of the intra-cluster medium. 

There is a known difference between the galaxy cluster temperatures measured by {\em XMM} and {\em Chandra}. This difference has been quantified with functions to calibrate the temperatures of one telescope to another; \cite{xmmchandracal} showed that the difference increases with temperature, with {\em XMM} EPIC temperatures being on average 7\% and 23\% lower than {\em Chandra} ACIS temperatures for 2~keV and 10~keV clusters respectively. Any analysis which uses both {\em XMM} and {\em Chandra} temperatures typically accounts for this \citep[e.g.,][]{farahixmmchandra,mikgasxmmchandra}. An understanding of whether there is a similar difference in {\em eROSITA} and {\em XMM} temperatures will be necessary before any joint analyses with data from the two telescopes are undertaken, and before scaling relations from one telescope can be safely used by another. 

In \S\ref{sec:efedsproperties} we explore the general properties of the eFEDS cluster catalogue and provide comparisons to a catalogue with similar properties. In \S\ref{sec:efedsxmm} we construct a cluster sample from the eFEDS catalogue with corresponding {\em XMM} observations, which includes a visual inspection of the X-ray data and SDSS/HSC images. We also compare eFEDS and XCS exposure times and central positions. In \S\ref{sec:meascomp} we compare luminosities and temperatures measured by eFEDS and XCS. Finally, in \S\ref{sec:discussion} we generate luminosity-temperature relations, discuss implications of our findings and how they can be improved. Then in \S\ref{sec:summary} we provide a final summary. The analysis code and samples are available in a GitHub repository\footnote{\href{https://github.com/DavidT3/eFEDS-XCS-Paper}{Analysis code and samples}}.

Throughout this work we use a concordance $\Lambda$CDM cosmology where $\Omega_{\rm{M}}$=0.3, $\Omega_{\Lambda}$=0.7, and $\rm{H}_{0}$=70 km s$^{-1}$ Mpc$^{-1}$, consistent with the original eFEDS cluster analysis (and other XCS works).

\section{Comparison of the \lowercase{e}FEDS Optically Confirmed and XXL-100-GC catalogues}
\label{sec:efedsproperties}

\begin{figure*}
    \centering
    \includegraphics[width=1.0\textwidth]{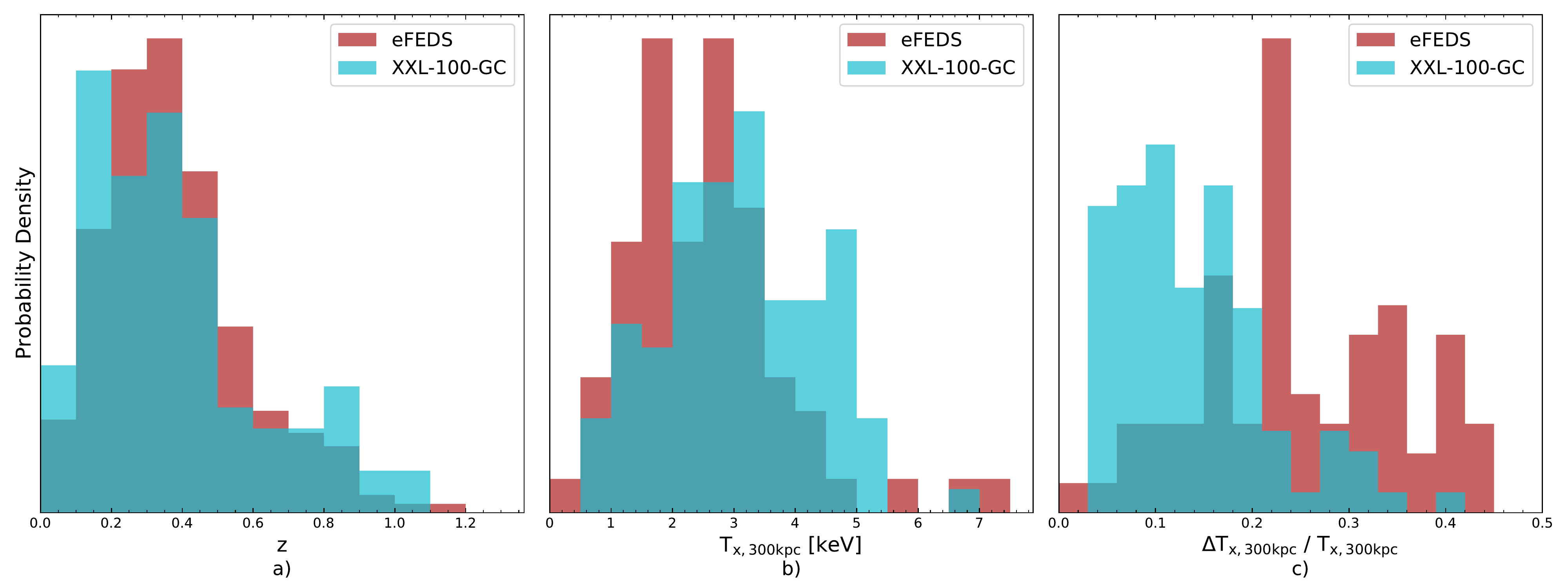}
    \caption[]{Redshift, temperature, and fractional temperature error distributions of the eFEDS and XXL-100-GC samples. Redshifts from both samples come from a variety of sources, and temperatures are measured within 300\,kpc apertures centered on clusters. There are no clusters in common between the two samples.} 
    \label{fig:xxlefeds}
\end{figure*}

The eFEDS cluster catalogue \citep{efedsclustercat} contains 542 candidates, 477 of which are considered to be optically confirmed \citep{efedsclusteropticalcat} when assessed using the Multi-component Matched Filter Cluster Confirmation Tool \citep[MCMF, ][]{MCMF}. All 542 X-ray candidates are accompanied by redshift ($z$) values. Soft-band (0.5-2.0~keV in the source frame) $L_{\rm X}$ values have been measured for 91\% of the X-ray cluster candidate sample. A smaller percentage, 21\%, of $T_{\rm X}$ values were obtained using spectra extracted from circular apertures centered on the eFEDS coordinates; 69 within 300\,kpc, and 95 within 500\,kpc (102 candidates have at least one temperature measurement).

The XXL survey \citep{xxlfoundation} covers ${\sim}50$~deg$^2$ of the sky (over two separate regions), making it the largest contiguous area survey in the {\em XMM} archive. It consists of 542 separate {\em XMM} observations with on-axis exposure times ranging from 10-20~ks. The contiguous nature of XXL makes it ideal to compare to eFEDS. As eFEDS and the XXL survey were taken in different parts of the sky there are no clusters in common. Although we note that the X-CLASS analysis of the {\em XMM} archive (up to August 2015) that made use of the XXL pipelines does contain some eFEDS candidates \citep[][]{xclass}. Comparisons are limited to ensemble distributions of the cluster samples.  We made use of the XXL-100-GC sample \citep{xxlgc100}, containing the 100 brightest galaxy clusters observed in XXL, and the sample of 477 optically confirmed eFEDS candidates.  The flux limits of the eFEDS and XXL-100-GC cluster samples are similar; ${\sim}10^{-14}\:\rm{erg}\:\rm{s}^{-1}\:\rm{cm}^{-2}$.

Figure~\ref{fig:xxlefeds}a shows the redshift distributions of the clusters in the two samples (eFEDS and XXL-100-GC distributions are shown in red and cyan respectively) to be very similar overall, but that XXL-100-GC detects a higher proportion of clusters at low redshifts.  Next, we compare the respective temperature distributions. Temperatures for the XXL-100-GC clusters were measured within a 300\,kpc aperture \citep{xxllt}, as were temperatures for eFEDS clusters, making a direct comparison of the distributions valid. Figure~\ref{fig:xxlefeds}b plots the temperature distributions of the two samples, with the XXL-100-GC temperature distribution containing a significantly higher proportion of temperatures above ${\sim}3.5\:\rm{keV}$. \cite{efedsclustercat} note that {\em eROSITA}'s ability to measure temperatures for hot clusters at $\gtrsim$5~keV is limited due to the reduced sensitivity of {\em eROSITA} at energies $>$3~keV. This is a plausible reason for the increased number of higher temperature clusters in XXL-100-GC compared to eFEDS. The effective area of {\em eROSITA} is ${\sim}$150~$\rm{cm}^2$ at 5~keV, compared to ${\sim}$900~$\rm{cm}^2$ for EPIC-PN, see Figure~9 in \cite{missionpaper} for a detailed comparison. Previous work by \cite{xcsmethod} has also shown that more counts are required to constrain temperatures to the same level for hotter galaxy clusters. Furthermore, the temperature distribution could also be influenced by the selection functions of the two surveys, differing measurement methodology, or a systematic difference in temperatures measured by the {\em eROSITA} and {\em XMM} telescopes (we explore this in Section~\ref{subsec:tcal}). 

\begin{figure*}
    \centering
    \includegraphics[width=1.0\textwidth]{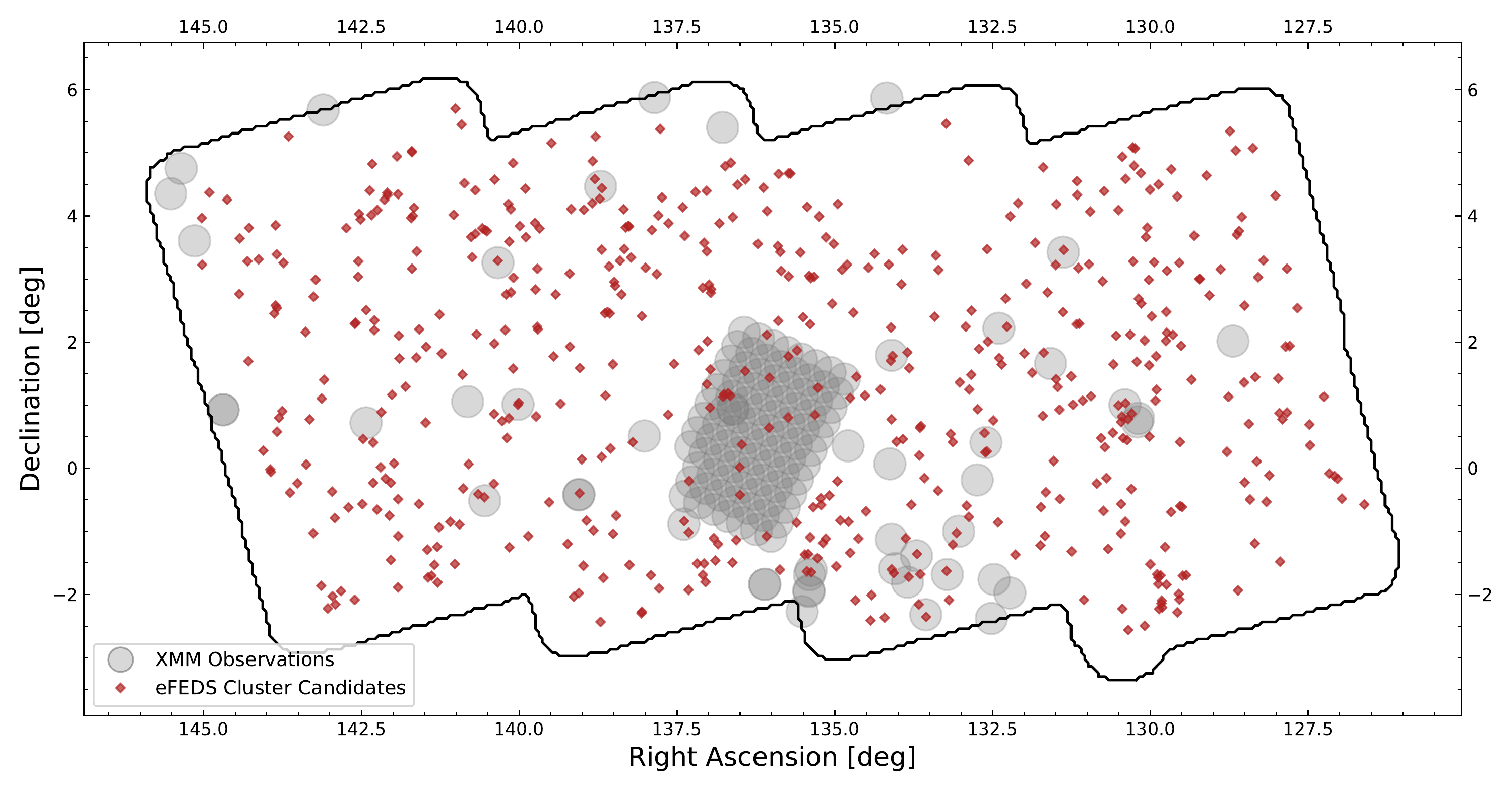}
    \caption[]{Footprint of eFEDS, given by the black solid line. Cluster candidates present in the eFEDS X-ray catalogue are highlighted by red diamonds. The grey circles highlight {\em XMM} observations, with a radius of 15\arcmin (the approximate radius of the {\em XMM} FoV).}
    \label{fig:efedsxcsclusters}
\end{figure*}

\begin{table*}
\begin{center}
\caption[]{{Summary of the samples defined in this work. N$_{\rm{cl}}$ is the number of clusters, N$_{T_{\rm{eFEDS, 500kpc}}}$ the number with eFEDS $T_{\rm{500kpc}}$ values, and N$_{T_{\rm{XCS, 500kpc}}}$ the number with XCS $T_{\rm{500kpc}}$ values.}\label{tab:samples}}
\vspace{1mm}
\begin{tabular}{l|lccc}
\hline
\hline
Sample Name & Description & N$_{\rm{cl}}$ & N$_{T_{\rm{eFEDS, 500kpc}}}$ & N$_{T_{\rm{XCS, 500kpc}}}$\\
\hline
\hline
eFEDS & The full eFEDS cluster candidates catalogue & 542 & 95 & - \\
\hline
eFEDS-{\em XMM} & eFEDS cluster candidates that fall on an {\em XMM} observation & 62 & 11 & - \\
\hline
eFEDS-XCS & eFEDS-{\em XMM} candidates available for analysis after inspection & 37 & 8 & 28 \\
\hline
\end{tabular}
\end{center}
\end{table*}

Finally we compare how well temperatures from the two samples are constrained, by comparing temperature uncertainties as a fraction of the absolute temperature value ($\Delta T_{\rm{X}}/T_{\rm{X}}$). Figure~\ref{fig:xxlefeds}c shows that, on average, XXL achieves better temperature constraints than eFEDS, with the mean percentage uncertainties for XXL and eFEDS being 14\% and 25\% respectively. This is consistent with the findings of \cite{xcsmethod}, who showed that ${\sim}1000$ background-subtracted soft-band (0.5-2.0~keV) counts are required to achieve a fractional temperature uncertainty of ${\sim}0.1$ for a 3~keV cluster. In this regard the longer exposures of XXL compared to eFEDS would give an advantage (especially in the deeper {\em XMM}-LSS fields, covering ${\sim}$11 deg$^{2}$ in the XXL-N field). 

\section{Understanding the \lowercase{e}FEDS catalogue contamination fraction}
\label{sec:efedsxmm}

In this section, we make use of archival {\em XMM} data that overlaps with the eFEDS footprint to assemble samples for analysis, and make an estimate of the contamination fraction in the eFEDS cluster candidate list. For this, we have used data products (images and source lists) generated by the {\em XMM} Cluster Survey \citep[XCS,][]{xcsfoundation}. The XCS source lists are constructed by the XCS Automated Pipeline Algorithm (XAPA) source finder, and a full explanation of our procedures can be found in \cite{xcsmethod}. We first determine which eFEDS X-ray cluster candidates fall within the active area of an {\em XMM} observation (Section~\ref{sec:eFEDS-XMM}). For these, we generate {\em eROSITA} cut-out images. We then compare, by eye, to the corresponding {\em XMM} cut-outs (Section~\ref{sec:efeds-verification}). We also compare to optical SDSS DR16 \citep[][]{dr16} images obtained from the SDSS cutout server\footnote{\href{https://skyserver.sdss.org/dr16/SkyserverWS/ImgCutout/getjpeg?ra=134.07450039749338&dec=-1.6358207558060298&width=1631&height=1631}{SDSS Image Cutout Server}}. As ${\sim}$70\% of the eFEDS-XMM sample are at redshifts $z$<0.5 it is generally appropriate to search for a red sequence using SDSS imagery, however for any candidate that we could not confirm with SDSS, we then examined images taken from the second data release of the Hyper Suprime-Cam Subaru Strategic Program \citep[HSC-SSP PDR2,][]{hsc_prd2}\footnote{\href{https://hsc-release.mtk.nao.ac.jp/das_cutout/pdr2/}{\color{blue}HSC-SSP PDR2 Image Cutout Server}}. The deeper data of HSC-SSP PDR2 ($i$-band limiting magnitude of 26.2 in the wide field, where the SDSS DR16 $i$-band limiting magnitude is 22.2) allow for detection of cluster galaxies at much higher redshifts.  Our visual inspection allows us to categorise contaminating objects (Section~\ref{sec:examplerejects}) and to estimate the overall contamination fraction in the eFEDS sample (Section~\ref{sec:contamfrac}).

\begin{figure}
    \centering
    \includegraphics[width=0.95\columnwidth]{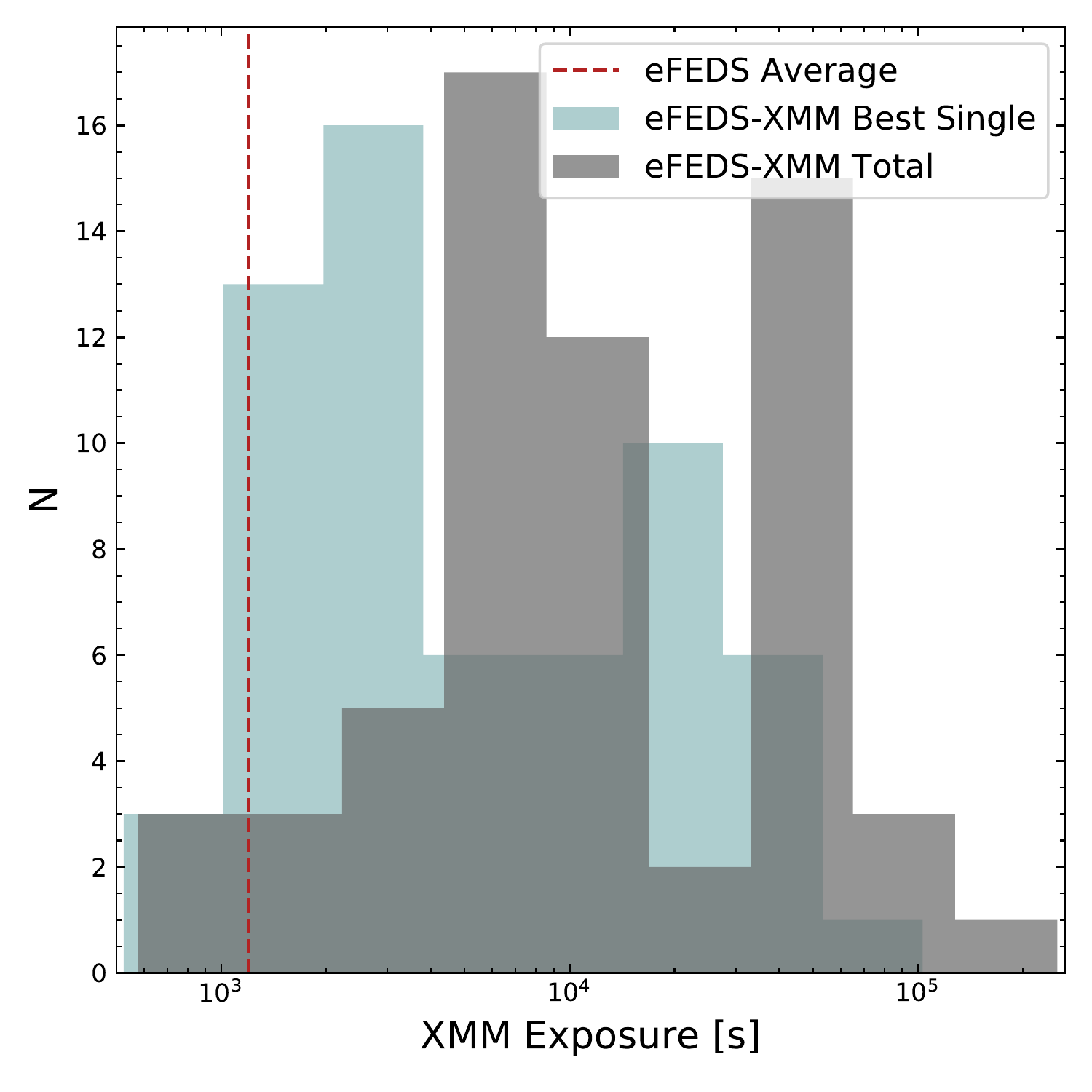}
    \caption[]{Distribution of exposure times for eFEDS-{\em XMM} cluster candidates, measured at the eFEDS coordinates. Exposures taken from 0.5-2.0~keV exposure maps, corrected for flaring and vignetting. Dashed line indicates the average vignetting corrected exposure of the eFEDS field reported by \cite{efedsclustercat}.}
    \label{fig:xmmexposure}
\end{figure}

\subsection{eFEDS Cluster Candidates in the {\em XMM} Footprint}
\label{sec:eFEDS-XMM}

\begin{figure}
    \centering
    \includegraphics[width=0.95\columnwidth]{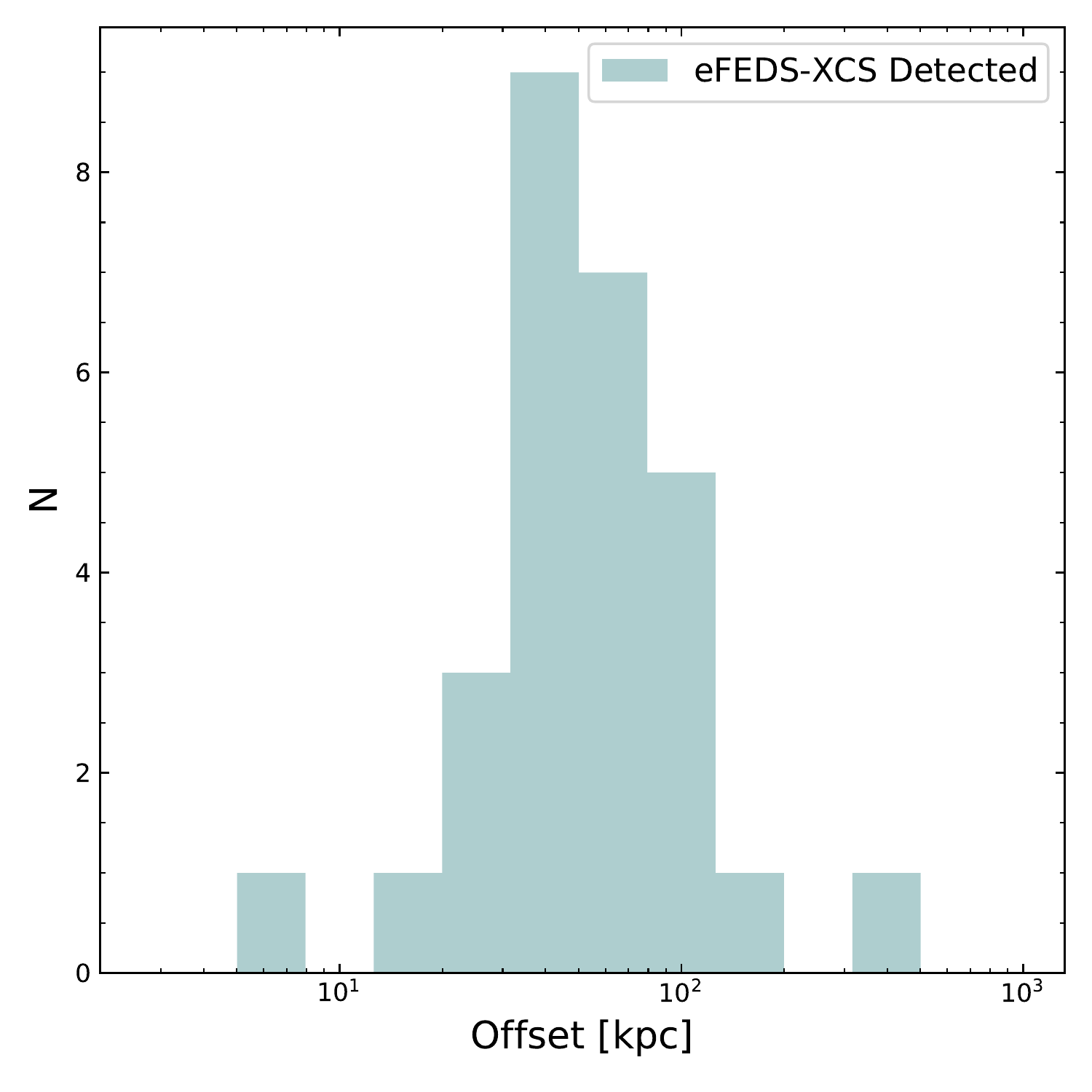}
    \caption[]{Comparison of eFEDS and XAPA central coordinates, for the subset of the eFEDS-XCS sample that have been detected by XAPA.}
    \label{fig:centcoords}
\end{figure}

\begin{figure*}
    \centering
    \includegraphics[width=1\textwidth]{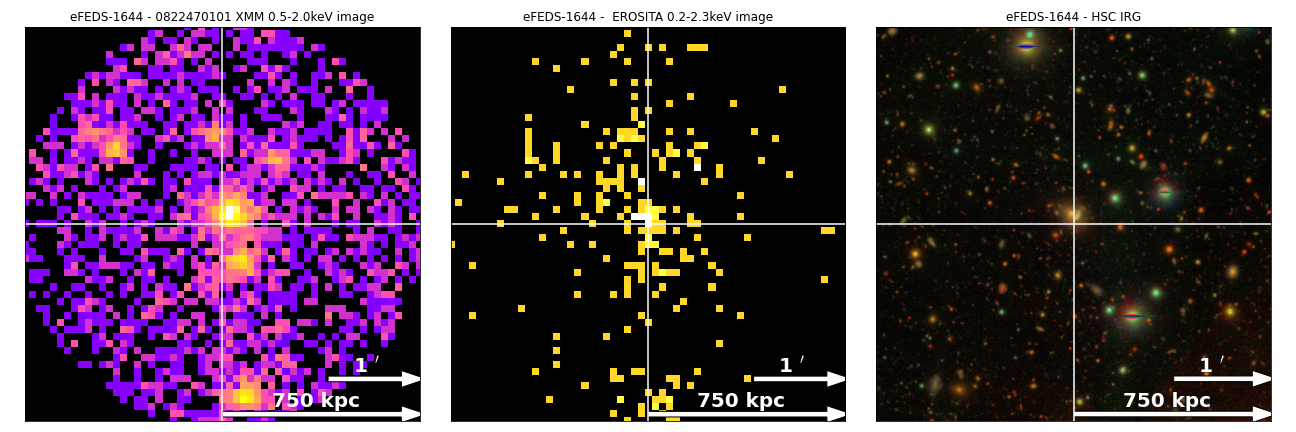}
    \caption[]{eFEDS-{\em XMM} cluster candidate (eFEDS ID 1644) identified as a pair interacting galaxies with ongoing AGN activity (see Section~\ref{subsubsec:blended}). The cross-hair indicates the eFEDS position. Left hand side is a combined PN+MOS1+MOS2 {\em XMM} image (ObsID 0822470101), centre is {\em eROSITA}, right hand side is HSC. Both {\em XMM} and {\em eROSITA} images are cutouts within a radius of 750~kpc, HSC image has a half-side-length of 750~kpc (at the redshift provided by eFEDS).}
    \label{fig:pairagn}
\end{figure*}

\begin{figure*}
    \centering
    \includegraphics[width=1\textwidth]{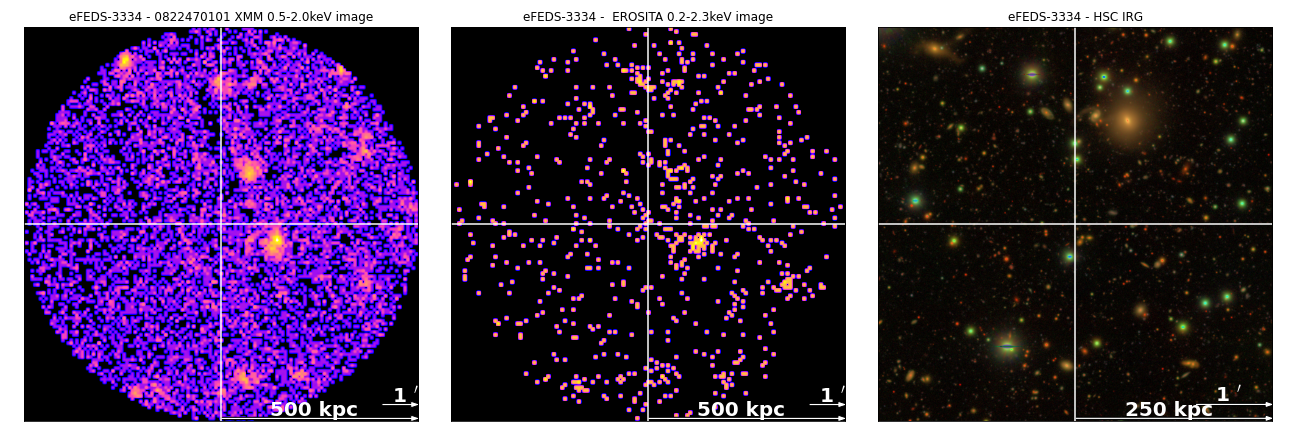}
    \caption[]{eFEDS-{\em XMM} cluster candidate (eFEDS ID 3334) without an obvious corresponding source of emission (see Section~\ref{subsubsec:spurious}). The cross-hair indicates the eFEDS position. Left hand side is a combined PN+MOS1+MOS2 {\em XMM} image (ObsID 0822470101), centre is {\em eROSITA}, right hand side is HSC. Both {\em XMM} and {\em eROSITA} images are cutouts within a radius of 500~kpc, HSC image has a smaller half-side-length of 250~kpc (at the redshift provided by eFEDS).}
    \label{fig:blanksky}
\end{figure*}

\begin{figure*}
    \centering
    \includegraphics[width=1\textwidth]{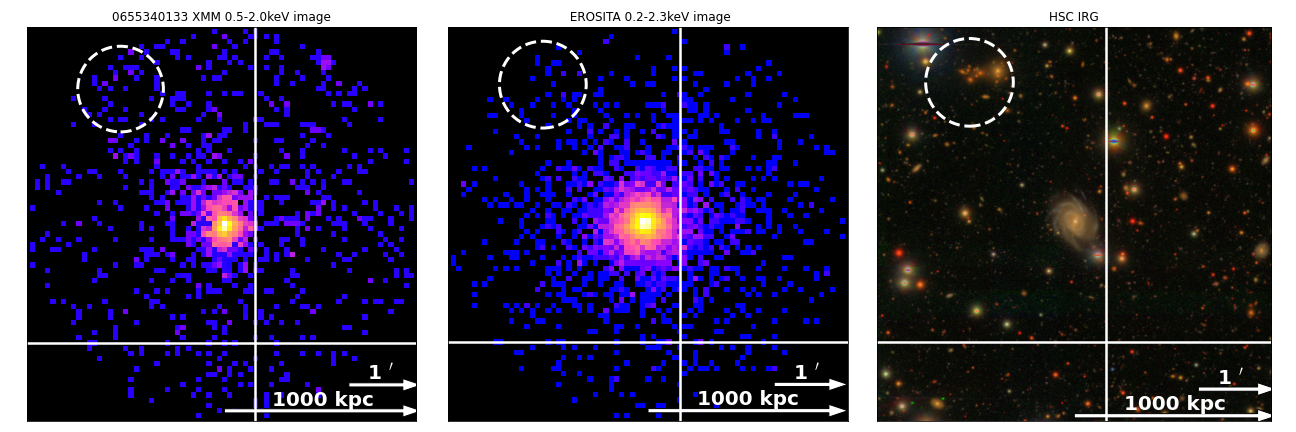}
    \caption[]{Two eFEDS-{\em XMM} cluster candidates in the outskirts of a low redshift foreground AGN. A spurious eFEDS-{\em XMM} cluster candidate (eFEDS ID 8922) is indicated by the cross-hair (see Section~\ref{subsubsec:spurious}). An eFEDS-{\em XMM} cluster candidate (eFEDS ID 16370) is indicated by the dashed circle. Left hand side is a combined PN+MOS1+MOS2 {\em XMM} image (ObsID 0655340133), centre is {\em eROSITA}, right hand side is HSC. Both {\em XMM} and {\em eROSITA} images are cutouts within a radius of 1000~kpc, HSC image has a half-side-length of 1000~kpc (at the redshift for eFEDS ID 8922 provided by eFEDS).}
    \label{fig:brightoutskirts}
\end{figure*}

\begin{figure*}
    \centering
    \includegraphics[width=1\textwidth]{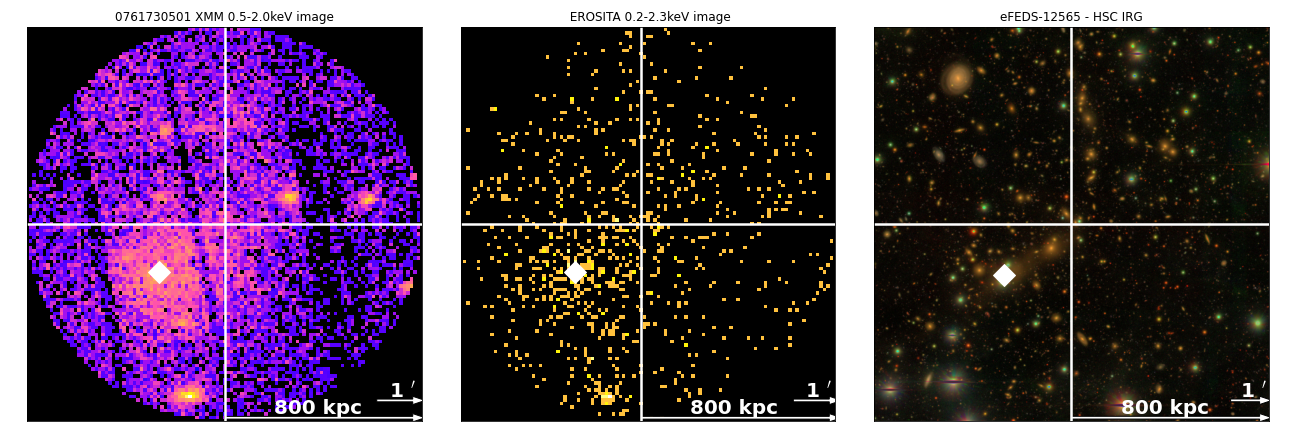}
    \caption[]{eFEDS galaxy cluster split into two candidates by the source finder (see Section~\ref{subsubsec:frag}). Cross-hairs indicate one candidate (eFEDS ID 8602), and the white diamond indicates the other (eFEDS ID 1023). Left hand side is a combined PN+MOS1+MOS2 {\em XMM} image (ObsID 0761730501), centre is {\em eROSITA}, right hand side is HSC. Both {\em XMM} and {\em eROSITA} images are cutouts within a radius of 800~kpc, HSC image has a half-side-length of 800~kpc (at the redshift provided by eFEDS, which is the same for 8602 and 1023).}
    \label{fig:splitcluster}
\end{figure*}

\begin{figure*}
    \centering
    \includegraphics[width=1\textwidth]{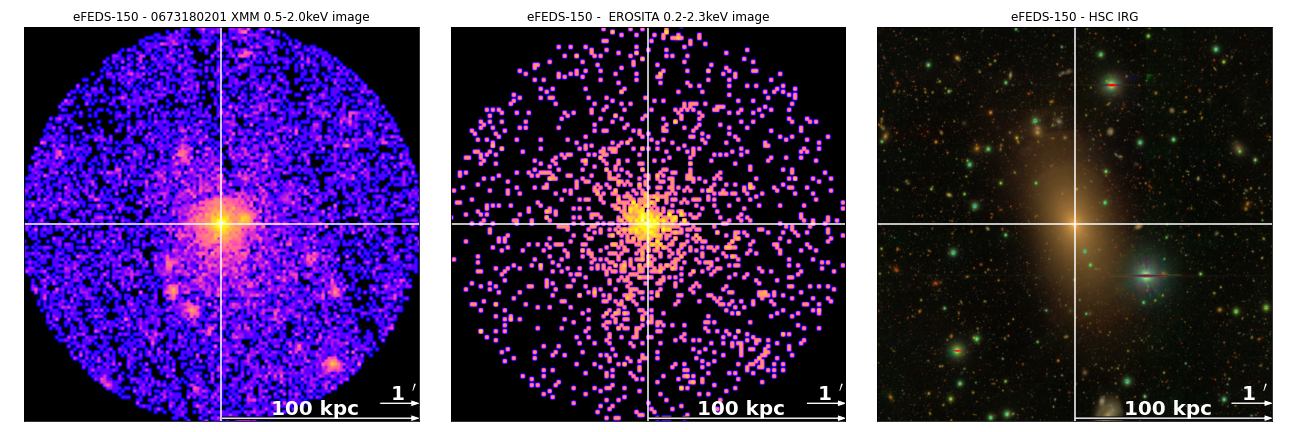}
    \caption[]{A low redshift eFEDS-{\em XMM} cluster candidate (eFEDS ID 150) whose X-ray emission is dominated by an X-ray bright elliptical galaxy (see Section~\ref{subsec:contamxray}). Left hand side is a combined PN+MOS1+MOS2 {\em XMM} image (ObsID 0673180201), centre is {\em eROSITA}, right hand side is HSC. Both {\em XMM} and {\em eROSITA} images are cutouts within a radius of 100~kpc, HSC image has a half-side-length of 100~kpc (at the redshift provided by eFEDS).}
    \label{fig:ellipticalxray}
\end{figure*}

Figure~\ref{fig:efedsxcsclusters} shows the outline of the eFEDS footprint with eFEDS X-ray cluster candidates \citep{efedsclustercat} indicated by red diamonds. {\em XMM} observations taken within the eFEDS footprint are indicated by grey shaded circles, with a radius of 15$^{\prime}$ (the approximate radius of the {\em XMM} field-of-view). There are a total of 143 {\em XMM} observations, covering ${\sim}$15\,deg$^{2}$ (11\%) of the sky within the eFEDS footprint, accounting for overlapping {\em XMM} observations.

We used {\em XMM}: Generate and Analyse \citep[\texttt{XGA}\footnote{\href{https://github.com/DavidT3/XGA}{{\em XMM}: Generate and Analyse GitHub}}][]{xgapaper}; a new, open-source, X-ray astronomy analysis module developed by XCS, to determine which of the 542 eFEDS cluster candidates listed in \cite{efedsclustercat} have also been observed by {\em XMM}.
An initial search finds eFEDS candidates with central coordinates within 30\arcmin of an {\em XMM} observation aimpoint (this is larger than the {\em XMM} field of view to account for any cases of low-$z$ clusters with centroid offsets). We then refined the match so that at least 70\% of a 300\,kpc aperture (centred on the eFEDS coordinate and assuming the eFEDS redshift) coincides with an {\em XMM} observation. Sixty-two eFEDS candidates met these criteria, and this sub-set is denoted the eFEDS-{\em XMM} sample (see Table~\ref{tab:samples}). Fifty-three of the eFEDS-{\em XMM} candidates appear in the eFEDS optically confirmed sample \citep[][]{efedsclusteropticalcat}.

The distribution of {\em XMM} exposure times for the eFEDS-{\em XMM} sample is shown in Figure~\ref{fig:xmmexposure}. The light-blue distribution uses the best individual observation exposure time for each eFEDS-{\em XMM} candidate; the grey distribution is the total exposure time for each candidate. These are vignetting-corrected exposure times at the eFEDS coordinate (rather than at the respective observation aim-point). The typical eFEDS vignetting-corrected exposure (1.2~ks) is shown by the dashed red line for comparison. The majority of exposure times (individual or total) are longer for {\em XMM} than eFEDS.

\subsection{Constructing the eFEDS-XCS Sample}
\label{sec:efeds-verification}

To judge the quality of the eFEDS candidates in the eFEDS-{\em XMM} sample, circular cut-out images, of radius 500\,kpc, were generated from the {\em eROSITA} eFEDS data and XCS processed {\em XMM} data. We select the {\em XMM} observation with the highest signal-to-noise within a 300~kpc aperture centered on the eFEDS candidate coordinate, and use its XCS generated PN+MOS1+MOS2 image \citep[][]{xcsgiles}. For the {\em eROSITA} cut-outs, the \texttt{eSASS}\footnote{\href{https://erosita.mpe.mpg.de/edr/DataAnalysis/}{Introduction to \texttt{eSASS}}} \texttt{evtool} software was used. 

Both sets of images used a pixel size of 4.35\arcsec, but different energy ranges were used; 0.5-2.0~keV for {\em XMM} and 0.2-2.3~keV for {\em eROSITA}. These ranges reflect those used by the respective XCS and eFEDS source detection routines.  

As shown in \cite{missionpaper}, the energy dependent effective area of {\em XMM} (assuming all three cameras are operating) is the same, or greater, than that of {\em eROSITA}. Therefore, we can expect most of the {\em XMM} cut-outs to have higher signal to noise than their eFEDS counterparts, even after accounting for the fact that the {\em XMM} background level is slightly higher than {\em eROSITA}'s. In 10 cases, however, we judged the {\em XMM} data to be inadequate for further analysis. This was either because the eFEDS candidate fell on the edge of the field of view and/or because the signal-to-noise was too low (see Table~\ref{tab:xmmrejects} for details). These 10 were excluded from further analysis, although it is noteworthy that in 5 cases, an obvious (by eye) extended source was visible in the corresponding {\em eROSITA} cut-out.

The remaining 52 eFEDS-{\em XMM} sources were then visually inspected side-by-side to judge the quality of the eFEDS candidates. Thirty-seven eFEDS-{\em XMM} candidates were confirmed as clusters suitable for X-ray analysis by this visual inspection, henceforth called the eFEDS-XCS sample. Generally speaking the XCS and eFEDS defined centroid positions were in good agreement (with an offset of less than 100~kpc), but several outliers are present (see Figure~\ref{fig:centcoords}). The outliers were due to either low signal to noise eFEDS data, or to eFEDS point source contamination. Similar examples were noted in \cite{efedsclusteropticalcat}. 

The other eFEDS cluster candidates were classified as sample contaminants (11, Table~\ref{tab:rejects}), or as having their X-ray flux contaminated by other sources (4, Table~\ref{tab:taintedxray}). There are three broad categories of sample contamination, as described below (Section~\ref{sec:examplerejects}).

\subsection{Categories of contaminating objects in the eFEDS X-ray cluster candidate catalogue}
\label{sec:examplerejects}
Here we discuss categories for the different types of contaminating object that we discovered in the eFEDS X-ray cluster candidate catalogue. The figures that we use to illustrate these examples are not necessarily on the same scale, or centered on the same position, as those we used for visual inspection, and all figures use HSC imagery for clarity. The optical images we used for general inspection were SDSS, with HSC photometry used in cases were we needed to clarify our classification.

\subsubsection{Blended sources}
\label{subsubsec:blended}

An example of this is eFEDS ID 1644, which is shown in Figure~\ref{fig:pairagn}; the two sources at the centre of the {\em XMM} image (Figure~\ref{fig:pairagn} left; detected as separate point sources by XCS) in the {\em XMM} cut-out appear as a single object in the {\em eROSITA} image. The dominant X-ray source is discussed in \cite{agnpair}. It is the result of AGN activity in a pair of interacting galaxies. The two galaxies can be seen in the corresponding HSC image in Figure~\ref{fig:pairagn} (right). The blending is likely a result of {\em eROSITA}'s 26{\arcsec} FOV average PSF half-energy width (HEW), which is larger than the {\em XMM} PN camera's 16.5{\arcsec} PSF HEW (at 1.5~keV), in combination with the short eFEDS exposure time. This source was assigned a class of B2 during the MCMF classification process, which indicates point source contamination, but it is still retained in the optically confirmed catalogue.

\subsubsection{Spurious sources}
\label{subsubsec:spurious}
An example of this is shown in Figure~\ref{fig:blanksky}. There does not appear to be a source at the eFEDS candidate centroid position (extended or otherwise) in either the {\em XMM} or {\em eROSITA} image (Figure~\ref{fig:blanksky}, left and middle respectively). We note that the redshift provided by the eFEDS cluster catalogue for this candidate (eFEDS ID 3334) is very low ($z{=}0.087$) and so any X-ray cluster emission should be obvious, unless it is very low surface brightness and/or very extended. Nonetheless, we do not see evidence of a coincident population of galaxies in the HSC imagery (Figure~\ref{fig:blanksky}, right). 

Another example is eFEDS ID 8922, which is shown in Figure~\ref{fig:brightoutskirts}; this cutout is not centered on the spurious eFEDS cluster candidate, but on the bright source that causes it. The eFEDS candidate location (at the cross hairs) is in the outskirts of a bright source (eFEDS ID 3, not present in the cluster candidate catalogue), and not coincident with any distinct source in either the {\em XMM} or {\em eROSITA} image (Figure~\ref{fig:brightoutskirts}, left and middle respectively). There also does not appear to be an association of galaxies in HSC imagery (Figure~\ref{fig:brightoutskirts}, right). The dominant X-ray source is identified as an AGN in the Million Quasar catalogue \citep{milliquas} located in a spiral galaxy visible in the corresponding optical image. It is also present in the eFEDS AGN catalogue \citep[][]{efedsagn}.

\subsubsection{Fragmented sources}
\label{subsubsec:frag}

An example of this is shown in Figure~\ref{fig:splitcluster}. The white cross-hair indicating the position of one eFEDS candidate (eFEDS ID 8602), and the white diamond another (eFEDS ID 1023). The two candidates have almost identical redshifts ($z{=}0.196$ and $z{=}0.197$ respectively). Luminosity measurements for both, and a temperature estimate for eFEDS ID 1023, are given in \cite{efedsclustercat}. We discuss this system further in Appendix~\ref{app:split_cluster}. We note that it is not used during our luminosity (Section~\ref{subsec:lumcomp}) and temperature comparisons (Section~\ref{subsec:tcomp} and Section~\ref{subsec:tcal}), or in our luminosity-temperature relation analysis (Section~\ref{subsec:LT}).

\subsection{Clusters with contaminated X-ray emission}
\label{subsec:contamxray}

In these cases there is evidence, from the SDSS and/or HSC data, for a physical association of galaxies -- which could in turn be responsible for an extended X-ray source due to  emission from a hot ICM  -- however, we contend that any ICM emission present is significantly contaminated by other X-ray sources. 
One example (eFEDS ID 150) is shown in Figure~\ref{fig:ellipticalxray}, where the emission detected by {\em eROSITA} appears to originate primarily from the central galaxy (alternative, but less likely explanations are that this is a fossil group or a system with a strong cool core). We note that similar examples were identified in eFEDS, candidates with IDs 3133 and 3008 (see Table~\ref{tab:taintedxray}).
An example of a different type of contaminated emission is presented in Figure~\ref{fig:brightoutskirts} (eFEDS ID 16370). The eFEDS candidate is highlighted by the white dashed circle. There is tentative evidence of X-ray emission in the {\em XMM} observation (especially when smoothing is applied), and the coordinates coincide with a collection of red galaxies in SDSS and HSC at an SDSS photo-$z$ of $z{=}0.44$ (matching the eFEDS catalogue's $z$). However, due to the proximity of the eFEDS candidate to the bright AGN (as discussed in Section~\ref{subsubsec:spurious}), the X-ray flux in this region will be contaminated by non ICM emission, so we exclude this cluster from the following analyses.

In this paper we focus on the comparison of the X-ray properties measured by {\em eROSITA} and {\em XMM}, so we do not include these four eFEDS candidates in the analyses presented Section~\ref{sec:meascomp}.
However, it would inappropriate to remove them from some other types of analyses -- such as cluster number count cosmology based on optical/near-IR selection  \cite{efedsclusteropticalcat} -- because they are still associated with galaxy over-densities. All of the sources in Table~\ref{tab:taintedxray} appear in the eFEDS optically confirmed sample.

\subsection{The eFEDS contamination fraction}
\label{sec:contamfrac}

As discussed in Section~\ref{sec:efeds-verification} (and collated in Appendix~\ref{app:rejected}), 11 of the 62 candidates (18\%, Table~\ref{tab:rejects}) in the eFEDS-{\em XMM} sample were not included in the eFEDS-XCS sample because they were classified  as being in one of the three sample contaminant types described in Section~\ref{sec:examplerejects}. This should be viewed as a lower limit because, in 10 (of 62) cases (Table~\ref{tab:xmmrejects}),  it was not possible for us to confirm the validity of the eFEDS candidate using archival {\em XMM} data. 

The eFEDS-XCS sample of eFEDS cluster candidates is an order of magnitude smaller than the full eFEDS X-ray cluster candidate catalogue.
Moreover, several of the eFEDS-XCS clusters were the target of their respective {\em XMM} observations, and that has been shown to introduce selection bias \citep[][]{xcsgiles}; this could influence {\em XMM} detections and thus the construction of our eFEDS-XCS sample. Even so, our result is consistent with simulations performed by the eRASS team that predicted a contamination level of $\sim$20\% \citep{simerass}. It is also consistent with the eFEDS optical counterparts study \citep[][]{efedsclusteropticalcat}, which measured a contamination fraction of 17$\pm 3$\%.

We also investigated whether any of the 11 cluster candidates that we classed as sample contaminants (Table~\ref{tab:rejects}) are present in the sample of 477 candidates that \cite{efedsclusteropticalcat} consider to be optically confirmed. We find that 5 of the 11 are present therein. This compares to 53 in the overall eFEDS-{\em XMM} sample, indicating a minimum contamination fraction of ${\sim}9\%$ in the \cite{efedsclusteropticalcat} sample. This is slightly high compared to the value of $6\pm 3$\% reported by \cite{efedsclusteropticalcat}. However, if we discount eFEDS ID 8602 and 3334 from consideration as sample contaminants \citep[to be more consistent with approach taken in][]{efedsclusteropticalcat}, the contamination level drops to ${\sim}$6\%. 

We note that one cluster of the eFEDS-XCS sample (eFEDS ID 5170) does not appear in the \cite{efedsclusteropticalcat} sample. This candidate was included in our sample because of its X-ray emission (in {\em XMM} and {\em eROSITA} images) and evidence of an over density of red galaxies in the SDSS and HSC photometry.

\begin{figure}
    \centering
    \includegraphics[width=0.95\columnwidth]{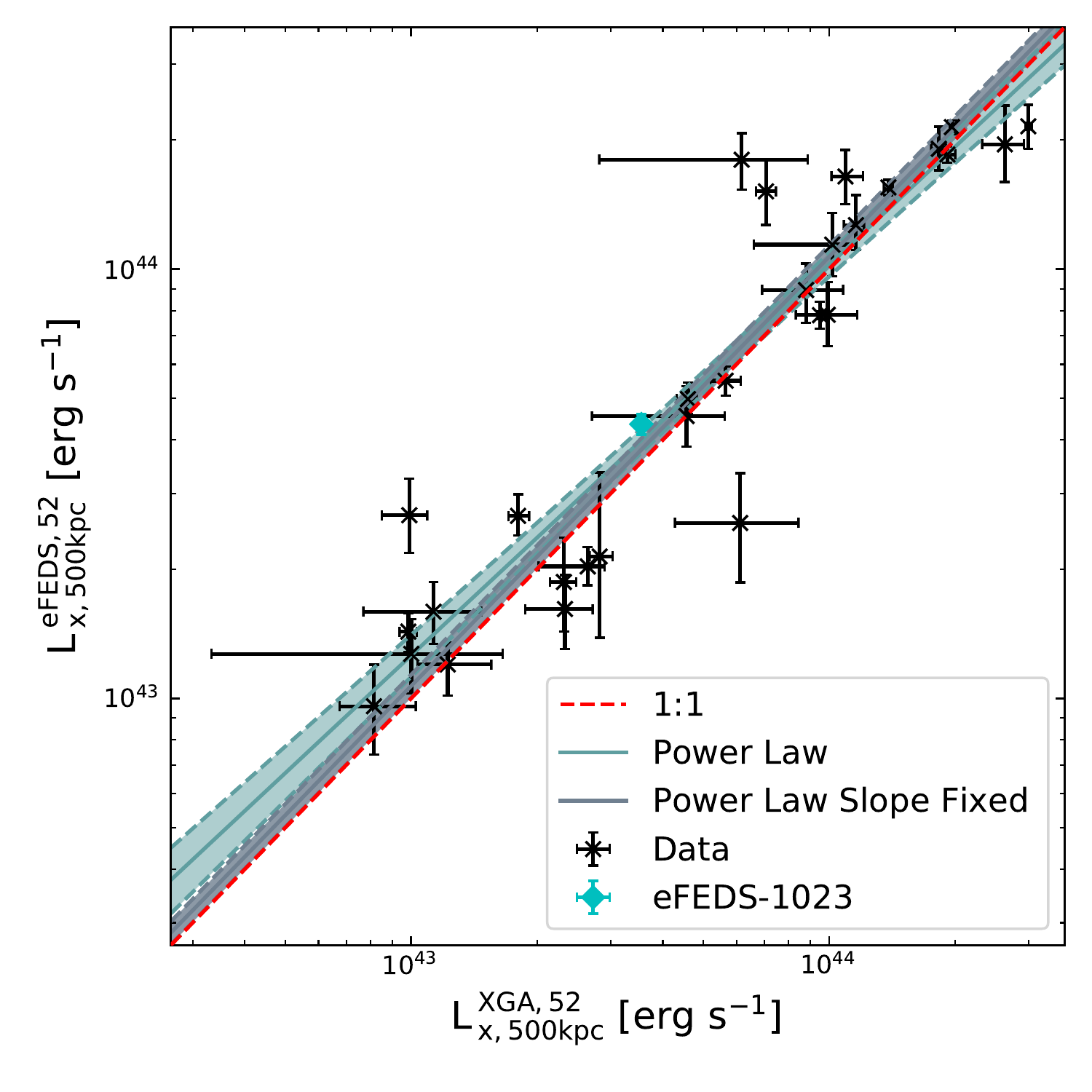}
    \caption[]{Comparison of unabsorbed cluster luminosities within a 500~kpc aperture, in the 0.5-2.0~keV energy band, centered on eFEDS coordinates. Pale blue line indicates best fit power-law, with 68\% confidence levels given by shaded region. Grey line indicates a power-law fit with slope set to 1 (with 68\% confidence levels given by grey shaded region). Cyan diamond is for the split cluster discussed in Appendix~\ref{app:split_cluster}.}
    \label{fig:l500kpccomp}
\end{figure}

\section{Comparisons of cluster properties measured by \lowercase{e}FEDS and XCS}
\label{sec:meascomp}

We use the \texttt{XGA} \citep[][]{xgapaper} XSPEC \citep[][]{xspec} interface to measure spectroscopic properties of those clusters that have high enough quality {\em XMM} data, and compare values to those presented in the eFEDS data release. Note that we do not re-analyse the eFEDS data, but compare to the measurements given by \cite{efedsclustercat}. We use \texttt{XGA} v0.2.1, SAS v17.0.0, and XSPEC v12.10.1.

\begin{figure*}
    \centering
    \includegraphics[width=1.0\textwidth]{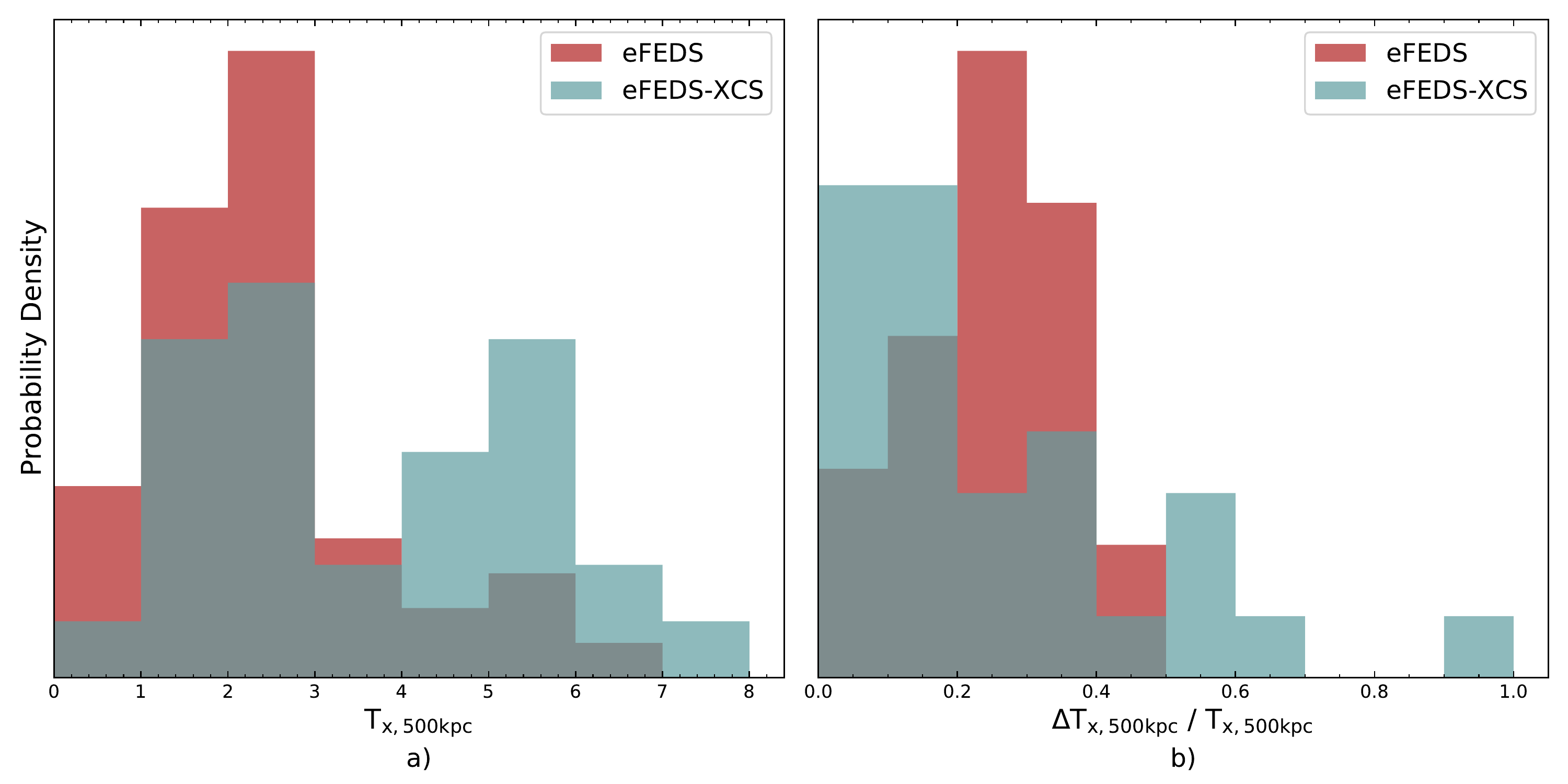}
    \caption[]{Temperature and fractional temperature error distributions of the eFEDS (red) and eFEDS-XCS (pale blue) samples, for measurements made within 500~kpc apertures, centered on eFEDS coordinates. } 
    \label{fig:efedsxmmtxdist}
\end{figure*}

\subsection{Fitting Procedure}
\label{subsec:fitproc}

Cluster spectra are extracted within a 500~kpc fixed aperture (as the eFEDS catalogue contains a greater number of 500~kpc temperatures than 300~kpc) and centered on the eFEDS position.  Corresponding backgrounds are extracted within 1000-1500~kpc annuli. Non-cluster sources in both the 500~kpc apertures and background regions are identified using the XCS region files, and their corresponding events are removed during spectrum generation.  We fit absorbed \cite[with \texttt{tbabs}, ][]{tbabs} plasma emission models \citep[APEC, ][]{apec} to the spectra; these models are standard for XCS analyses, but are also the same as those used in the eFEDS spectroscopic analysis. To maximise the similarity of our analysis to eFEDS we opt to use the abundance tables published by \cite{aspl} when performing our spectral fits. The abundance parameter of the APEC model in all cases is frozen at 0.3~Z$_{\odot}$, the nH parameter of the \texttt{tbabs} model is set from the full-sky HI survey by the \cite{nh} using the HEASoft \texttt{nh} tool and frozen. The redshift parameter is set to the eFEDS catalogue value and frozen. The temperature is initially set to 3~keV and the normalisation is initially set to 1~cm$^{-5}$, then both are allowed to vary.

Each spectrum is first fit independently, and if the measured temperature is outside of the range $0.01~{\rm keV} < T_{\rm{X}} < $20~keV, or either temperature uncertainty is $>$ 15~keV, then the spectrum will not be included in the final fit. A simultaneous fit is then performed using only the spectra that fulfil the requirements outlined above. Temperatures and unabsorbed luminosities are then determined from this joint fit. We used a given temperature measurement in further analyses (Sections~\ref{subsec:tcomp},  ~\ref{subsec:tcal}, and \ref{subsec:LT}) if, a) the best fit value is less than 25~keV, b) the upper and lower uncertainties are both positive, and c) the larger uncertainty is less than three times the smaller. Likewise, fitted luminosities are used in our analyses (Sections~\ref{subsec:lumcomp} and \ref{subsec:LT}), if both the upper and lower uncertainties are not greater than the best fit value, and if the upper and lower uncertainties are both positive. 

For a more complete explanation of the spectral fitting process and comparisons of results with other {\em XMM} analyses that confirm the veracity of measurements produced by this procedure, see \cite{xcsmassmethod}. All measurements for the eFEDS-XCS sample can be found in Table~\ref{tab:measurements}, along with eFEDS ID, position, and redshift.

\subsection{Luminosity Comparison}
\label{subsec:lumcomp}

We compare luminosities measured with both {\em XMM} and {\em eROSITA}, since one of the main products of eRASS will be large catalogues of X-ray cluster luminosities. These will be used as the basis of various {\em eROSITA} science applications; for example, a mass-luminosity scaling relation \citep[such as the one recently produced by][]{efedsmor} provides a way to estimate masses and overdensity radii of a given cluster, enabling X-ray cluster cosmology. Therefore, it is important to test the fidelity of eFEDS luminosities with {\em XMM} data. 

The eFEDS analysis presents cluster luminosities measured via a forward-fitting analysis of 2D count-rate maps, including considerations of the morphology of the cluster, rather than by the fitting of emission models to spectroscopic data. In the context of eFEDS, this allows for the measurement of accurate luminosities for clusters that do not have high enough quality data to perform spectral fitting. We can directly compare {\em XMM} and {\em eROSITA} luminosities for 29 (${\sim}$80\%) of the eFEDS-XCS sample. We use a spectral fitting process to measure unabsorbed (corrected for hydrogen column absorption) luminosities in the soft (0.5-2.0keV) energy band for those clusters with a successful {\em XMM} temperature measurement.

We fit a power-law with the slope fixed at unity and another power-law with the slope left to vary to the luminosity comparison, finding the results of both to be entirely consistent with a one-to-one relation. The fits were performed in log space using the R package LInear Regression in Astronomy\citep[{\sc lira}\footnote{\href{https://cran.r-project.org/web/packages/lira/index.html}{LInear Regression in Astronomy}}, ][]{softlira}, fully described in \cite{LIRA}. Figure~\ref{fig:l500kpccomp} demonstrates an excellent soft-band luminosity agreement (including the two models) between eFEDS and XCS, especially considering the differing measurement methods. We also include the data point for the cluster eFEDS ID 1023 (discussed further in Appendix~\ref{app:split_cluster}), but do not include it in our comparison fit. Luminosities measured by eFEDS and XCS are similarly well constrained, though the XCS uncertainties tend to be slightly smaller.

\subsection{Temperature Comparison}
\label{subsec:tcomp}

We have been able to measure temperatures within a 500 kpc aperture for ${\sim}$80\% of the eFEDS-XCS sample, though only ${\sim}$30\% of those also have an eFEDS temperature available (see Table~\ref{tab:samples} for a summary). We first compare the overall temperature, and fractional temperature uncertainty, distributions, as we did in Section~\ref{sec:efedsproperties} with the XXL-100-GC sample.

Figure~\ref{fig:efedsxmmtxdist}a, which shows the overall temperature distributions of the eFEDS and eFEDS-XCS samples, demonstrates that a larger proportion of {\em XMM} temperatures than {\em eROSITA} temperatures are above ${\sim}$4~keV, similar to the behaviour in Figure~\ref{fig:xxlefeds}b with the XXL-100-GC sample. It is likely that this is due to the difference in telescope sensitivity at high energies, as well as other selection effects resulting from targeted {\em XMM} exposures. 

Figure~\ref{fig:efedsxmmtxdist}b demonstrates that a larger proportion of the eFEDS-XCS clusters (compared to eFEDS measurements) have a fractional temperature uncertainty of less than 20\%, and as such the {\em XMM} temperature measurements are generally better constrained. However we also note that the temperature fractional error distribution of the eFEDS-XCS sample extends to larger values than eFEDS. 

In summary, archival {\em XMM} observations can provide temperatures which are, on average, better constrained than eFEDS for those clusters that have been observed by {\em XMM}, and can also deliver more temperatures for hotter systems due to {\em XMM}'s greater sensitivity at high energies.  As such, the {\em XMM} archive will be a very useful complement to the eRASS.

\begin{figure}
    \centering
    \includegraphics[width=0.95\columnwidth]{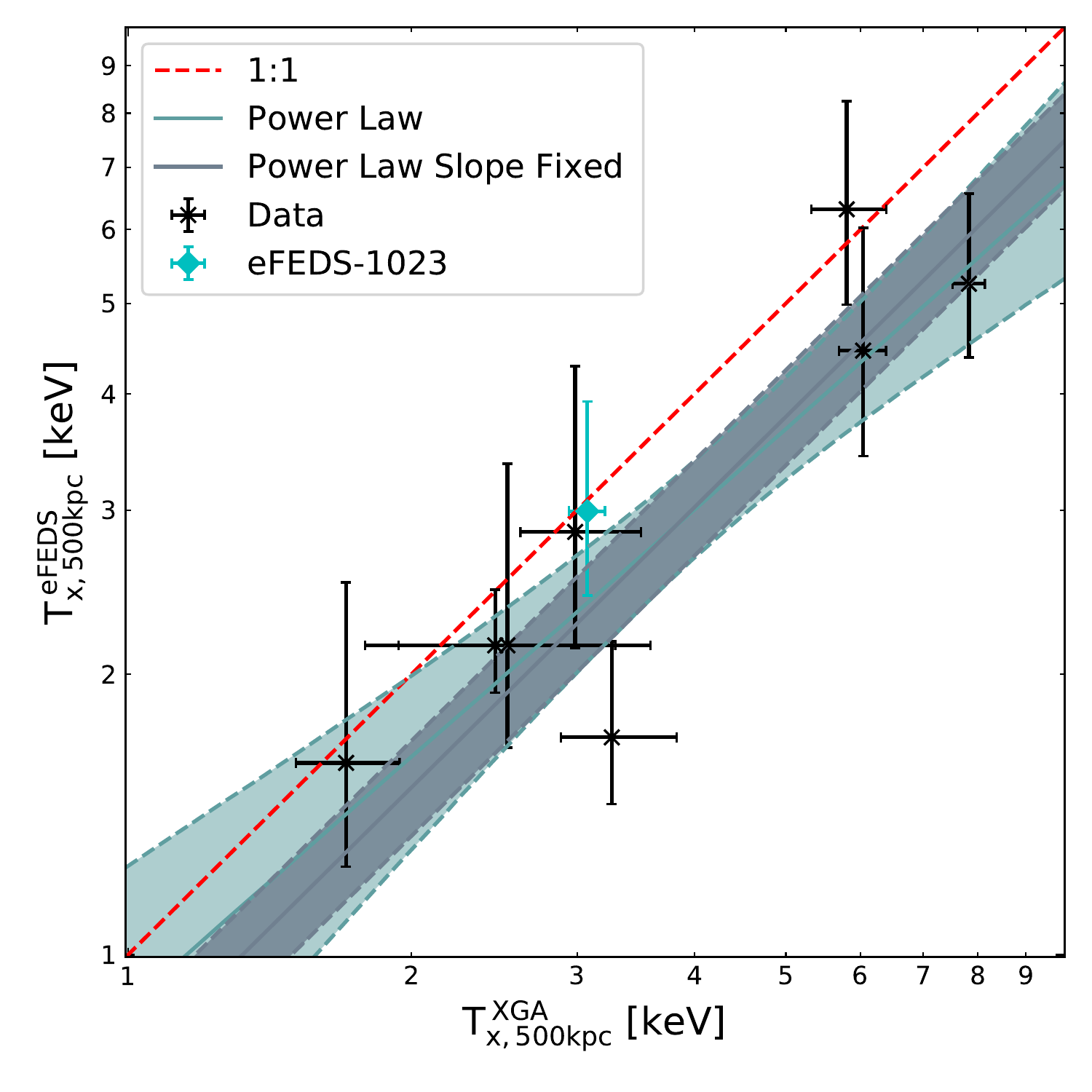}
    \caption[]{Comparison of eFEDS and XCS cluster temperatures within 500~kpc, centered on eFEDS coordinates. Pale blue line indicates best fit power-law (slope free to vary), with 68\% confidence levels given by shaded region. Grey line indicates a power-law fit with fixed slope of unity (with 68\% confidence levels given by the grey shaded region). Cyan diamond is for the split cluster discussed in Appendix~\ref{app:split_cluster}.}
    \label{fig:t500kpccomp}
\end{figure}

\begin{table}
\begin{center}
\caption[]{{\small The normalisation, slope, and intrinsic scatter values of the fitted temperature calibration models for 500~kpc apertures. $A_{TT}$ is normalisation, $B_{TT}$ is slope, and $\sigma_{T_{\rm{eROSITA}}|T_{\rm{XMM}}}$ the intrinsic scatter.}\label{tab:tempcal}}
\vspace{1mm}
\begin{tabular}{cccc}
\hline
\hline
Calibration Name & $A_{TT}$ & $B_{TT}$ & $\sigma_{T_{\rm{eROSITA}}|T_{\rm{XMM}}}$\\
\hline
\hline
Power Law & $0.88^{+0.37}_{-0.29}$ & $0.89^{+0.25}_{-0.24}$ & $0.04^{+0.06}_{-0.03}$ \\
\hline
Power Law Fixed Slope & $0.75^{+0.10}_{-0.08}$ & 1 & $0.04^{+0.05}_{-0.02}$ \\
\hline
\end{tabular}
\end{center}
\end{table}

\subsection{Temperature Calibration}
\label{subsec:tcal}
Motivated by the known temperature offset between {\em XMM} and {\em Chandra} \citep[][]{xmmchandracal}, we test for a difference in temperatures measured by {\em XMM} and {\em eROSITA}. We make a comparison between temperatures presented in the eFEDS cluster catalogue and {\em XMM} temperatures that we have measured for the same clusters. 

The comparison of 8 clusters with measured {\em XMM} and {\em eROSITA} (eFEDS) temperatures is given in Figure~\ref{fig:t500kpccomp}, and shows a systematic offset between the two telescopes. We also plot the data point for the cluster with eFEDS ID 1023 (discussed further in Appendix~\ref{app:split_cluster}), but do not include it in our comparison fit. All but one of the clusters are below the one-to-one line, indicating that the {\em eROSITA} temperatures are systematically lower than their {\em XMM} counterparts. To model this, we fit a power law of the form

\begin{align}
{\rm log}\left(T^{\:\rm{eROSITA}}_{\rm{x, 500kpc}}\right) &= {\rm log}(A_{TT}) + B_{TT}{\rm log}\left(T^{\:\rm{XMM}}_{\rm{x, 500kpc}}\right) \pm \sigma_{T_{\rm{eROSITA}}|T_{\rm{XMM}}},
\label{equ:TvsT}
\end{align}
where $A_{TT}$ is the normalisation, $B_{TT}$ the slope and $\sigma_{T_{\rm{eROSITA}}|T_{\rm{XMM}}}$ the intrinsic scatter. The fits were performed in the same way as those in Section~\ref{subsec:lumcomp}. The best fit parameters are given in Table~\ref{tab:tempcal}. 

First, we fit the power law with the slope left free to vary, probing whether the observed offset evolves with temperature \citep[as found in][comparing between {\em Chandra} and {\em XMM}]{xmmchandracal}. We measure a slope value of 0.89$^{+0.25}_{-0.24}$, indicating that the calibration evolves with system temperature.  We note however that due to the large errors, the value of $B_{TT}$ is consistent with 1 (within 1$\sigma$).  The measured intrinsic scatter of both fits is very low (essentially consistent with zero), which is as expected. Due to the large errors on the measured slope, we re-fit the power-law with the slope fixed at unity. This allows us to measure a single overall normalisation that describes the average difference in temperatures measured by the two telescopes.  We measure a normalisation of 0.75$^{+0.10}_{-0.08}$, meaning that (on average) {\em eROSITA} measures a temperature ${\sim}25\%$ cooler than those measured by {\em XMM} for the same cluster. 

As we have not re-analysed {\em eROSITA} data and measured our own {\em eROSITA} temperatures for the eFEDS-XCS cluster sample, the observed temperature calibration offset could be the result of some mismatch in our respective methodologies. The accuracy of the measured offset is also limited by small number statistics, as very few clusters have both an {\em eROSITA} and an {\em XMM} temperature. However, we have provided evidence of an offset that requires further investigation.

\section{Discussion}
\label{sec:discussion}
In this work we have presented a measurement of the temperature offset between {\em eROSITA} and {\em XMM} for a sample of galaxy clusters.  Here we discuss potential impacts of this offset on the derived scaling relations and how the temperature calibration can be improved.  

\subsection{Comparison of {\em eROSITA} and {\em XMM} X-ray scaling relations}
\begin{figure}
    \centering
    \includegraphics[width=1\columnwidth]{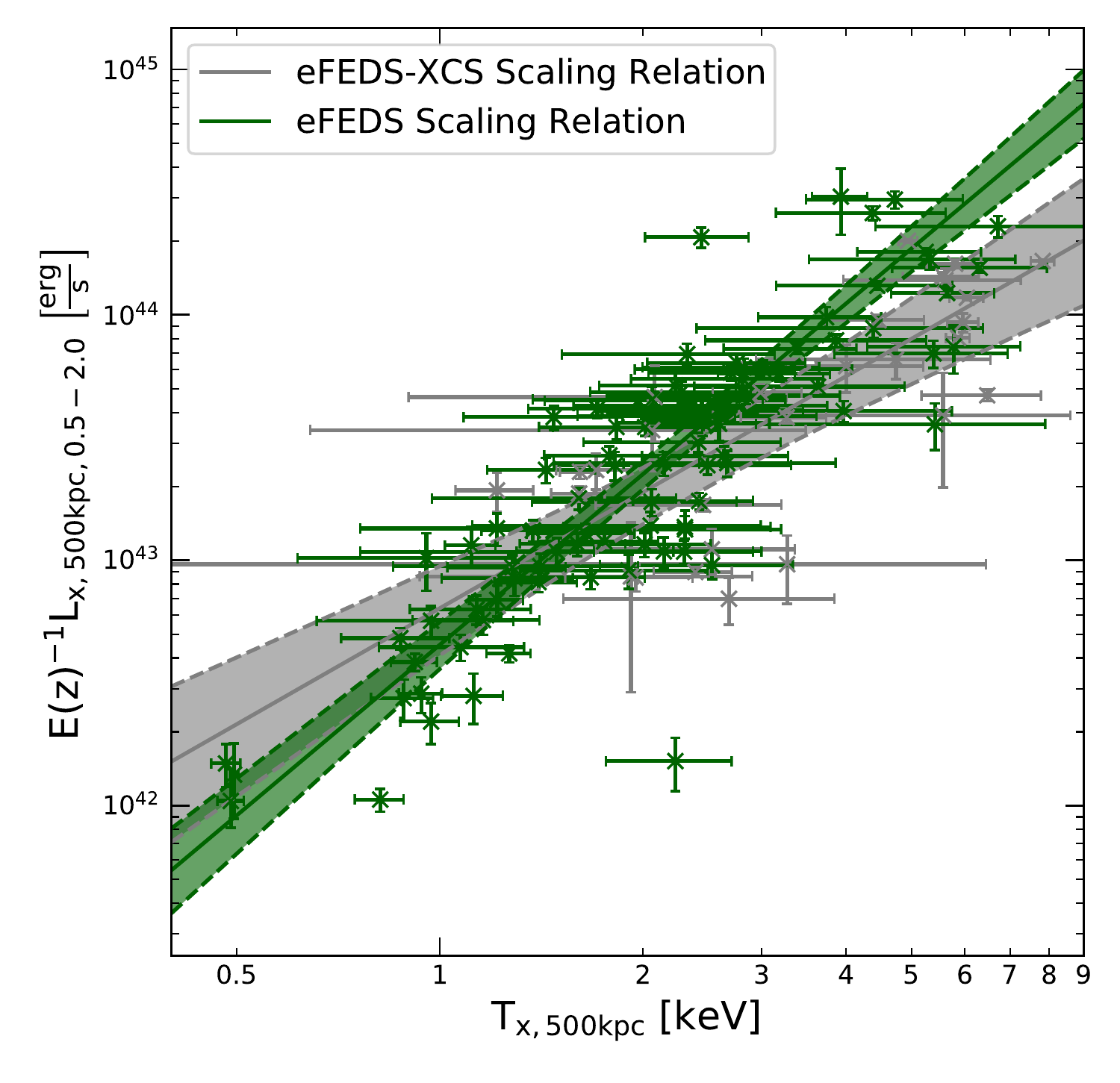}
    \caption[]{Soft-band (0.5-2.0~keV) luminosity-temperature relations for the eFEDS and eFEDS-XCS samples. Properties measured within a 500~kpc fixed aperture centered on the eFEDS positions. eFEDS data points are green, eFEDS-XCS data points are grey.}
    \label{fig:efedsandxcslt}
\end{figure}

\begin{figure}
    \centering
    \includegraphics[width=1\columnwidth]{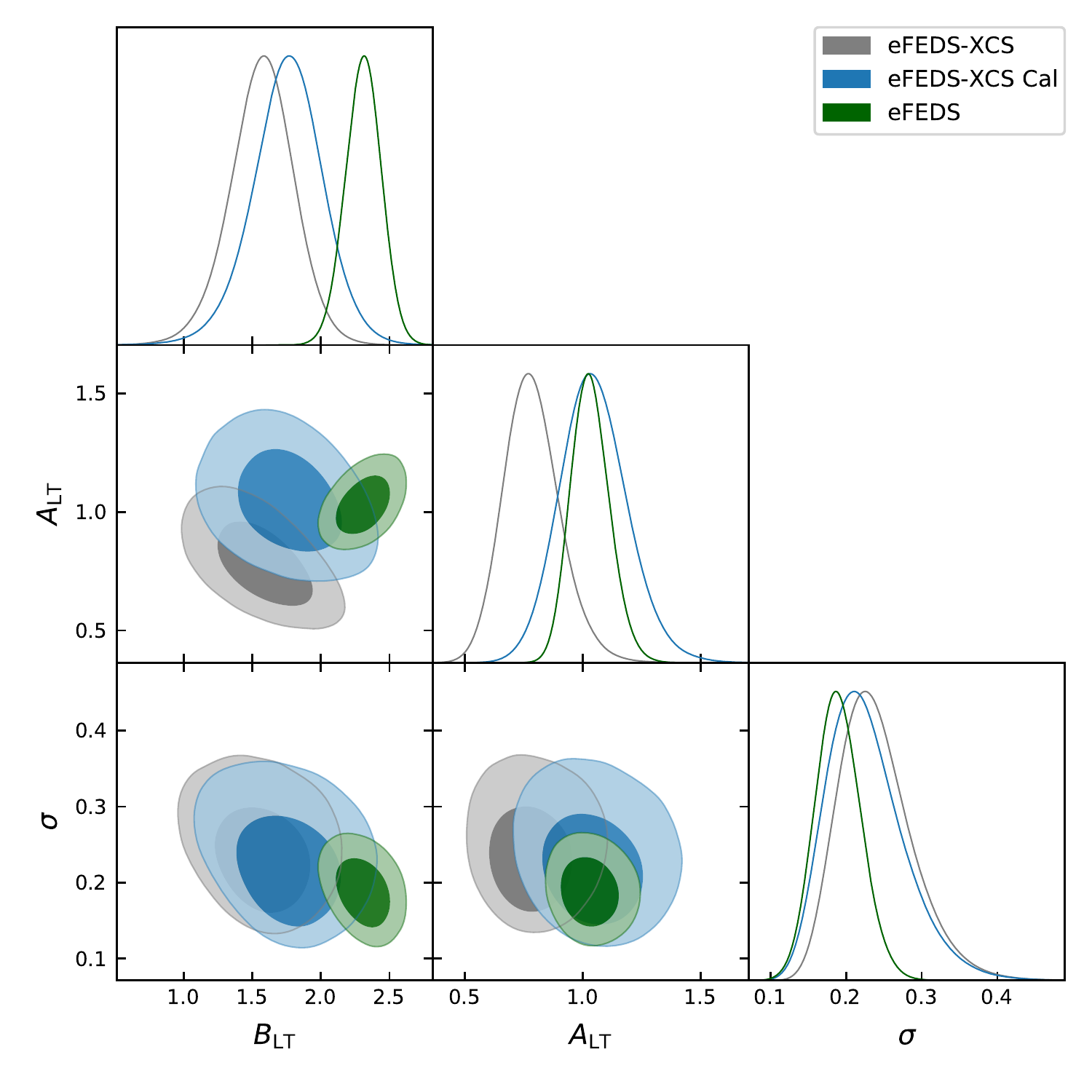}
    \caption[]{Corner plot of the 1$\sigma$ and 2$\sigma$ confidence contours of the $L^{500\rm{kpc}}_{\rm{x, 0.5-2.0}}$ - $T^{500\rm{kpc}}$ relation parameters, for the eFEDS (green contours), eFEDS-XCS (grey contours) and eFEDS-XCS calibrated (blue contours) samples.  The diagonal shows the posterior densities of each parameter.}
    \label{fig:ltcorner}
\end{figure}

\begin{figure}
    \centering
    \includegraphics[width=1\columnwidth]{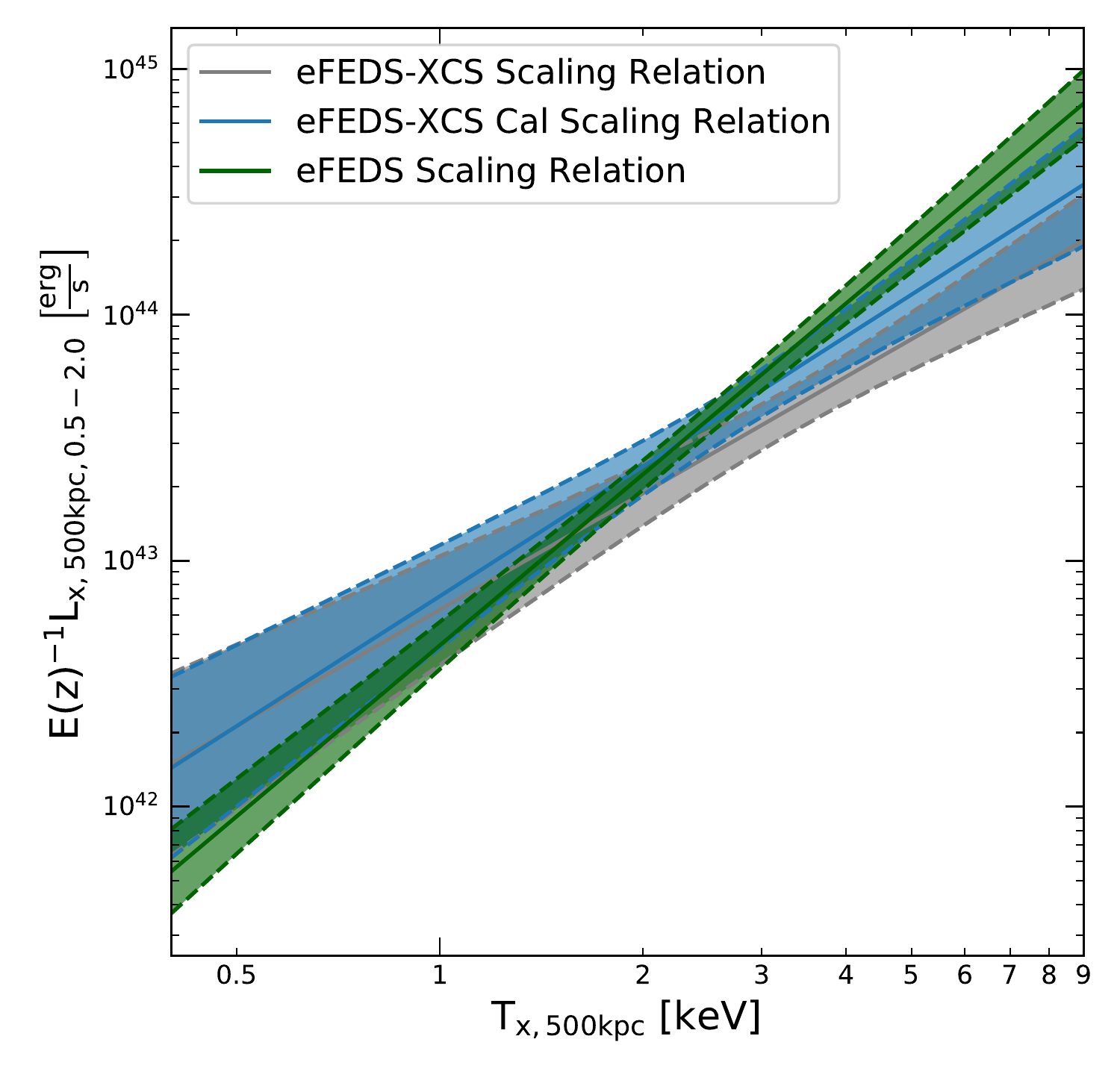}
    \caption[]{Soft-band luminosity-temperature relations for eFEDS, eFEDS-XCS, and calibrated eFEDS-XCS. Properties measured within a 500~kpc fixed aperture centered on the eFEDS positions.}
    \label{fig:prelimlt}
\end{figure}

\begin{table}
\begin{center}
\caption[]{{\small The normalisation, slope, and residual scatter values from the LIRA fits of the different datasets, for the $L^{500\rm{kpc}}_{\rm{X, 0.5-2.0}}$ - $T_{\rm{X}}^{500\rm{kpc}}$ scaling relation.}\label{tab:relations}}
\vspace{1mm}
\begin{tabular}{cccc}
\hline
\hline
Relation Name & $A_{LT}$ & $B_{LT}$ & $\sigma_{L|T}$\\
\hline
\hline
eFEDS & $1.03^{+0.09}_{-0.08}$ & $2.31^{+0.13}_{-0.13}$ & $0.19^{+0.03}_{-0.03}$\\
\hline 
eFEDS-XCS & $0.78^{+0.12}_{-0.11}$ & $1.58^{+0.22}_{-0.23}$ & $0.23^{+0.05}_{-0.04}$\\
\hline 
eFEDS-XCS Calibrated & $1.04^{+0.15}_{-0.14}$ & $1.76^{+0.25}_{-0.26}$ & $0.22^{+0.06}_{-0.04}$\\
\hline
\end{tabular}
\end{center}
\end{table}

\label{subsec:LT}
We have explored the impact on X-ray scaling relations in light of the temperature offset measured in Section~\ref{subsec:tcal}.  We focus on the luminosity-temperature relation derived from eFEDS and XCS data.  
We use 28 eFEDS-XCS clusters with a successful 500~kpc temperature ($T^{\rm 500kpc}_{\rm X}$) and soft-band luminosity ($L^{\rm 500kpc}_{X,0.5-2.0}$) measurement (instead of limiting the analysis to the 8 clusters used in Section~\ref{subsec:tcal} for temperature calibration), and all available eFEDS candidates and compare the relations.  

The $L^{500\rm{kpc}}_{\rm{X,0.5-2.0}}$ - $L_{\rm{X}}^{500\rm{kpc}}$ relations for eFEDS (grey points) and eFEDS-XCS (green points) are shown in Figure~\ref{fig:efedsandxcslt}; the eFEDS relation uses data from 94 clusters.
We fit both sets of data using a power law of the form, 
\begin{align}
{\rm log}\left(\frac{L^{500\rm{kpc}}_{\rm{X,0.5-2.0}}}{E(z)L_{0}}\right) &= {\rm log}(A_{LT}) + B_{LT}{\rm log}\left(\frac{T^{\rm{500kpc}}_{\rm{X}}}{T_{0}}\right) \pm \sigma_{L|T},
\label{equ:lt}
\end{align}
where $A_{LT}$ denotes the normalisation, $B_{LT}$ the slope, and $\sigma_{L|T}$ the intrinsic scatter of the relation. We calculate $E(z)$ using the redshift supplied in the eFEDS catalogue and our chosen concordance cosmology. The fits are performed using the LIRA package. We set normalisation values for luminosity and temperature to approximate median eFEDS values;  $L_{0}$=3.0$\times$10$^{43}$~erg~s$^{-1}$ and $T_{0}$=2.3~keV.

Figure~\ref{fig:efedsandxcslt} shows the best-fit relations  using the eFEDS (green line) and eFEDS-XCS (grey line) samples respectively.  The best-fit values are given in Table~\ref{tab:relations}, with their distributions illustrated in Figure~\ref{fig:ltcorner}.  While the distributions highlight that the parameters of the relation are consistent within their 2$\sigma$ contours, the difference in the central value of the slope warrants further discussion.

\begin{figure*}
    \centering
    \includegraphics[width=0.98\textwidth]{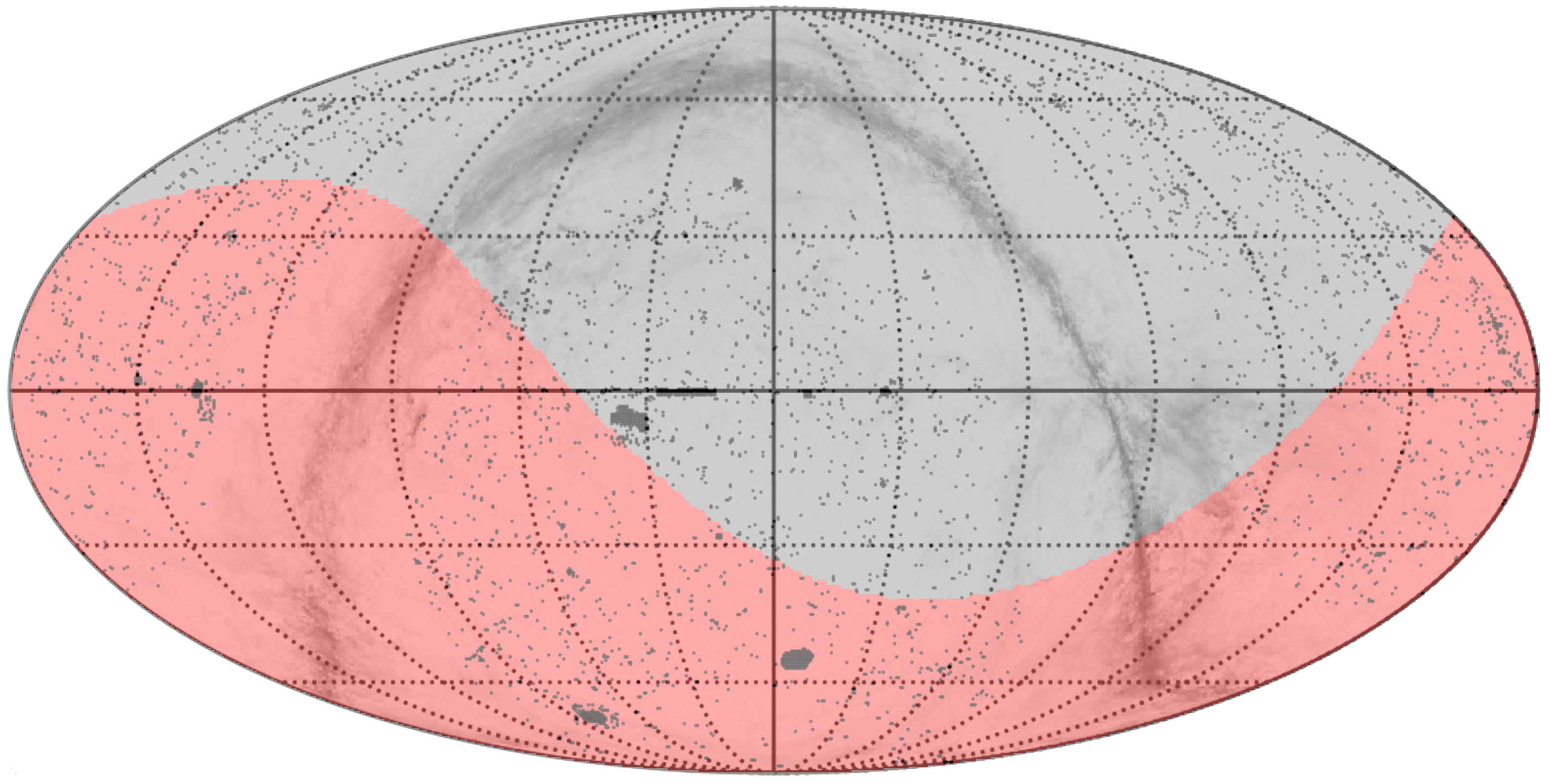}
    \caption[]{Distribution of {\em XMM} observations projected over the sky, indicated by grey points. The eRASS-DE half of the sky is highlighted by the red region. The background is the Planck-DustPol \citep[][]{dustpol} map, available for download on NASA's \href{http://lambda.gsfc.nasa.gov/data/footprint-maps/Planck_DustPol_Amp_256.fits.gz
}{Lambda service}.} 
    \label{fig:germanerassxmm}
\end{figure*}

We explore whether this difference can be reduced by accounting for the observed temperature offset found in Section~\ref{subsec:tcal}. We measure a third version of the scaling relation, designed to test the effect of the temperature calibration quantified in Section \ref{subsec:tcal}.  We determine a ``calibrated'' luminosity-temperature relation by using the power law model (with slope free to vary), with parameter values provided in Table~\ref{tab:tempcal}, to convert the eFEDS-XCS {\em XMM} temperature values to predicted {\em eROSITA} values.

Figure~\ref{fig:prelimlt} shows the model fits for all three relations (with data points omitted for clarity), and shows that the eFEDS-XCS calibrated scaling relation has a steepened slope and increased normalisation when compared to the original eFEDS-XCS relation.  Figure~\ref{fig:ltcorner} shows a shift of the contours and distributions (the eFEDS-XCS calibrated parameters are given by the blue contours) towards eFEDS (blue contours).  The normalisation of the calibrated eFEDS-XCS relation is fully consistent with the eFEDS relation, with the tension in the measured slopes reduced.  

\subsection{Future work to improve the calibration of the {\em XMM} to {\em eROSITA} temperature offset}

The measured temperature discrepancy described in Section~\ref{subsec:tcal} and shown in Figure~\ref{fig:t500kpccomp} is based on only 8 sets of measurements, of which only 3 are more than one sigma from the one-to-one relation. Further investigation is required to quantify the level of a temperature offset between the {\em XMM} and {\em eROSITA}. For this, we plan three complementary approaches:
\begin{itemize}
\item{Re-analyse the eFEDS cluster candidate observations using an identical spectroscopic methodology to the {\em XMM} analysis. This way, we will have control of all aspects of the analysis, including which regions are used to mask the observations for spectrum generation (see a related discussion in Appendix~\ref{app:split_cluster}). This will be done for all 94 clusters in the eFEDS luminosity-temperature analysis (Figure~\ref{fig:efedsandxcslt}), which includes the eight clusters featured in Figure~\ref{fig:t500kpccomp}.}
\item{Propose {\em XMM} follow-up observations of a representative sample of eFEDS clusters with robustly (i.e. percentage error less than 25\%) measured {\em eROSITA} temperatures. There are 43 such examples in the eFEDS data release that are not already included in the analysis shown in Figure~\ref{fig:t500kpccomp}. We will preferentially select clusters that fill gaps at the high and low temperature ends, to better constrain a temperature dependent slope in the calibration relation (if one exists).}
\item{Repeat the analysis herein after the next {\em eROSITA}-DE data release (due in Q4 2022\footnote{\href{https://erosita.mpe.mpg.de/erass/}{{\em eROSITA}-DE Data Release Schedule}}). This will cover the whole Southern sky (red area in Figure~\ref{fig:germanerassxmm}) and thus overlap with many more archival {\em XMM} observations than did eFEDS (grey points and regions in Figure~\ref{fig:germanerassxmm}). The next data release will have an exposure time eight times less shorter than eFEDS (one pass), but, with ${\sim}$150 times the area, we can still expect it to yield roughly 1000 robust temperature measurements.}
\end{itemize}

\section{Summary}
\label{sec:summary}

In this work we have performed the first comparison between cluster properties measured by the eFEDS survey and those measured by {\em XMM} surveys, both directly and for ensembles of clusters. A comparison of XXL-100-GC and the eFEDS optically confirmed sample indicated that the two samples have very similar redshift distributions, that XXL-100-GC contained proportionally more temperature measurements above 3.5~keV, and that XXL-100-GC temperatures are, on average, better constrained than eFEDS temperatures (14\% vs 25\% average percentage uncertainties respectively).

We have located and analysed eFEDS cluster candidates that have a counterpart in an {\em XMM} observation; as part of this process we visually inspected eFEDS cluster candidates that are within an {\em XMM} observation and rejected any that had no ICM emission at the eFEDS coordinates, had obviously contaminated ICM emission, or too low quality {\em XMM} data. The eFEDS-{\em XMM} sample contains 62 eFEDS cluster candidates that have been observed by {\em XMM}, and the eFEDS-XCS sub-sample contains the 37 clusters that pass our visual inspection (Table~\ref{tab:samples}). During visual inspection we found that 10 candidates (Table~\ref{tab:xmmrejects}) did not have sufficient {\em XMM} data to confirm or deny a cluster, 4 had X-ray flux contamination (Section~\ref{subsec:contamxray}, Table~\ref{tab:taintedxray}), and 11 were sample contaminants (Section~\ref{sec:examplerejects}, Table~\ref{tab:rejects}). We found that the majority of eFEDS-{\em XMM} candidates had a longer exposure time (at their eFEDS position) in {\em XMM} than in eFEDS. We also found that the majority of eFEDS centroid positions for the eFEDS-XCS sample are within 100~kpc (at the eFEDS redshift) of the XCS centroid positions (for clusters with an XCS match), though some outliers exist due to low signal-to-noise eFEDS data.

Our visual inspections of the eFEDS cluster candidates that fall on an {\em XMM} field (using {\em eROSITA}, {\em XMM}, SDSS and HCS images) have shown that there are some aspects of {\em eROSITA}'s source finding and confirmation steps that can introduce spurious sources into their catalogues, which in turn could impact their cosmological analyses. Our inspection process finds that the eFEDS-{\em XMM} sample is (at minimum) ${\sim}$18\% contaminated, and that optically confirmed sample is ${\sim}$9\% (${\sim}$6\% if made more consistent with the eFEDS sample selection method) contaminated. This is consistent with predictions from simulations \citep[][]{simerass} and eFEDS measured values \citep[][]{efedsclusteropticalcat}.

We have presented comparisons between cluster luminosities and temperatures measured with {\em eROSITA} and {\em XMM}. Our analysis finds excellent agreement between soft-band luminosities measured by eFEDS and XCS for the eFEDS-XCS sample (Section~\ref{subsec:lumcomp}), which is very encouraging for future eRASS cosmology analyses. Such analyses will rely almost exclusively on mass-luminosity relations \citep[such as the recent eFEDS-HSC work, ][]{efedsmor}, as even full depth eRASS will not be able to measure temperatures for enough galaxy clusters to use mass-temperature relations for cosmology. 
An ensemble cluster comparison was performed (Section~\ref{subsec:tcomp}) between the whole eFEDS cluster candidate sample (that had successful temperature/luminosity measurements), and the eFEDS-XCS sample (that had successful temperature/luminosity measurements). It showed similar results to the XXL-100-GC comparison, with a greater fraction of {\em XMM} temperature measurements being above 4~keV than eFEDS temperatures, and better average temperature constraints for {\em XMM} measurements.

A discrepancy between cluster temperatures measured by eFEDS and XCS has been found and quantified (Section~\ref{subsec:tcal}), with {\em eROSITA} temperatures being (on average) $25{\pm}9$\% cooler than those measured by {\em XMM} for the same cluster. This could hint at the need for a calibration function between the {\em eROSITA} and {\em XMM} telescopes (as has been necessary between {\em XMM} and {\em Chandra}). Several variables need to be better controlled before we can definitively state that the observed discrepancy is entirely due to a required temperature calibration, but we consider it to be a likely explanation. 

We also fit and compare luminosity-temperature scaling relations using data from the eFEDS catalogue and {\em XMM} data for the eFEDS-XCS sample, using them to compare scaling relations measured with {\em eROSITA} and {\em XMM} data. This has particular relevance for anyone wishing to use an {\em XMM} generated scaling relation in an {\em eROSITA} analysis, or vice versa, as we find a distinct tension between scaling relations from the eFEDS and eFEDS-XCS samples. We also generate a second, calibrated, eFEDS-XCS scaling relation which is in better agreement with the eFEDS relation; normalisation becomes entirely consistent, and the slope measurement tension is reduced. This again may suggest that a temperature scaling between {\em XMM} and {\em eROSITA} is necessary. It is likely, however, that a large part of the observed tension between the scaling relations is due to selection effects.

Our comparisons have shown that we can expect a great deal of useful data from the full eRASS catalogues, and that {\em XMM}-Newton still has a significant part to play as followup instrument. Its archive of 20 years worth of observations is still extremely valuable to the {\em eROSITA} team, and the X-ray astronomy community as a whole, and there will be many excellent opportunities for synergies between the two telescopes.

\section*{Acknowledgements}
This work is based on data from eROSITA, the soft X-ray instrument aboard SRG, a joint Russian-German science mission supported by the Russian Space Agency (Roskosmos), in the interests of the Russian Academy of Sciences represented by its Space Research Institute (IKI), and the Deutsches Zentrum für Luft- und Raumfahrt (DLR). The SRG spacecraft was built by Lavochkin Association (NPOL) and its subcontractors, and is operated by NPOL with support from the Max Planck Institute for Extraterrestrial Physics (MPE). The development and construction of the eROSITA X-ray instrument was led by MPE, with contributions from the Dr. Karl Remeis Observatory Bamberg \& ECAP (FAU Erlangen-Nuernberg), the University of Hamburg Observatory, the Leibniz Institute for Astrophysics Potsdam (AIP), and the Institute for Astronomy and Astrophysics of the University of Tübingen, with the support of DLR and the Max Planck Society. The Argelander Institute for Astronomy of the University of Bonn and the Ludwig Maximilians Universität Munich also participated in the science preparation for eROSITA.

The eROSITA data shown here were processed using the eSASS software system developed by the German eROSITA consortium.

We made use of TOPCAT \citep[][]{topcat} during the initial exploration of the eRASS cluster catalogue.

The new X-ray analysis module developed by XCS (\texttt{XGA}) makes significant use of Astropy \citep[][]{astropy1, astropy2}, NumPy \citep[][]{numpy}, Matplotlib \citep[][]{matplotlib}, and pandas \citep[][]{pandassoftware,pandaspaper}. XGA also uses GetDist \citep[][]{getdist} to produce corner plots.

DT, KR, and PG acknowledge support from the UK Science and Technology Facilities Council via grants ST/P006760/1 (DT),  ST/P000525/1 and ST/T000473/1 (PG, KR). PTPV was supported by Fundação para a Ciência e a Tecnologia (FCT) through research grants UIDB/04434/2020 and UIDP/04434/2020.
\section*{Data Availability}

The data underlying this article were accessed from the \href{https://erosita.mpe.mpg.de/edr/eROSITAObservations/Catalogues/}{eROSITA Early Data Release site}, and the {\em XMM} science archive. The derived data underlying this article are available in the article and in its online supplementary material.



\bibliographystyle{mnras}
\bibliography{xcs_efeds} 




\appendix

\section{Excluded Cluster Candidates}
\label{app:rejected}

\begin{table*}
\begin{center}
\caption[]{{eFEDS-{\em XMM} galaxy cluster candidates excluded from further analysis due to one or more {\em XMM}-Newton data quality issues.\newline $^\dagger$ indicates that the candidate was present in the optically confirmed sample from \cite{efedsclusteropticalcat}.}\label{tab:xmmrejects}}
\vspace{-5mm}
\begin{tabular}{cccc|p{0.70\linewidth}}
\hline
\hline
eFEDS ID & RA & Dec & $z$ & Notes\\
\hline
\hline
8094$^\dagger$ & 133.644 & -1.677 & 0.595 & On the edge of the {\em XMM} field of view. \\ 
\hline
7700 & 133.669 & -2.159 & 0.472 & On the edge of the {\em XMM} field of view. \\ 
\hline
1797$^\dagger$ & 133.876 & -1.11 & 0.754 & On the edge of the {\em XMM} field of view (the {\em eROSITA} image confirms presence of an extended source). \\ 
\hline
11836$^\dagger$ & 135.272 & -1.424 & 0.405 & On the edge of the {\em XMM} field of view, and the {\em XMM} observation is shallow (1023 s exposure at the eFEDS coordinates). The {\em eROSITA} image confirms presence of an extended source. \\ 
\hline
9877$^\dagger$ & 136.04 & 0.642 & 0.311 & Low signal-to-noise {\em XMM} image (7489 s exposure at the eFEDS coordinates) which has been affected by flaring.  \\ 
\hline
2757$^\dagger$ & 134.756 & 1.114 & 0.162 & Low signal-to-noise {\em XMM} image (6870 s exposure at the eFEDS coordinates). \\ 
\hline
5858 & 136.687 & 1.19 & 0.441 & Low signal-to-noise {\em XMM} image (15020 s exposure at the eFEDS coordinates). {\em eROSITA} image confirms presence of extended source. \\ 
\hline
2074$^\dagger$ & 136.971 & 1.569 & 0.163 & Low signal-to-noise {\em XMM} image (5331 s exposure at the eFEDS coordinates). \\ 
\hline
11837$^\dagger$ & 138.201 & 0.413 & 0.308 & Low signal-to-noise {\em XMM} image (29479 s exposure at the eFEDS coordinates) which has been affected by flaring. The {\em eROSITA} image confirms presence of an extended source. \\ 
\hline
1376$^\dagger$ & 133.23 & -1.627 & 0.343 & {\em XMM} data too shallow for confirmation. The {\em eROSITA} and SDSS data indicate a likely cluster, but eFEDS coordinate is offset from the extended emission and SDSS galaxies.  \\
\hline
\end{tabular}
\end{center}
\end{table*}

This section details the eFEDS-{\em XMM} X-ray cluster candidates that were not included in eFEDS-XCS sample, as discussed in Section \ref{sec:efedsxmm}. Basic information about the samples used in this work is available in Table~\ref{tab:samples}. Table~\ref{tab:xmmrejects} contains candidates that were not included due to the low quality of the {\em XMM} data available, Table~\ref{tab:taintedxray} contains galaxy clusters whose X-ray emission has been significantly contaminated by another X-ray source and as such were not included in the eFEDS-XCS sample. Table~\ref{tab:rejects} contains sample contaminants that were not included in the eFEDS-XCS sample (see Section~\ref{sec:efeds-verification}).

\begin{table*}
\begin{center}
\caption[]{{eFEDS-{\em XMM} galaxy cluster candidates classed as contaminants during our visual inspection of {\em XMM}, {\em eROSITA}, and SDSS images.\newline $^\dagger$ indicates that the candidate was present in the optically confirmed sample from \cite{efedsclusteropticalcat}.}\label{tab:rejects}}
\vspace{-5mm}
\begin{tabular}{cccc|p{0.70\linewidth}}
\hline
\hline
eFEDS ID & RA & Dec & $z$ &  Notes\\
\hline
\hline
1644$^\dagger$ & 130.396 & 1.031 & 0.507 & Blend: In {\em XMM} image two point sources are detected, due to {\em XMM}'s smaller PSF effect. The source is the target of the {\em XMM} observation and is associated with an interacting pair of active galaxies. (see Figure~\ref{fig:pairagn}).\\ 
\hline
3334$^\dagger$ & 130.508 & 0.995 & 0.087 & Spurious: There is not an X-ray source at this location in either the {\em XMM} or {\em eROSITA} images, nor do there appear to be any associated galaxies in the SDSS/HSC images. \\ 
\hline
8602$^\dagger$ & 132.593 & 0.269 & 0.196 & Fragmented: The ICM emission from a single cluster that has been classified as coming from two eFEDS candidates, ID 8602 and 1023. \\ 
\hline
5909$^\dagger$ & 133.83 & -1.721 & 0.365 & Spurious: There is an X-ray source at the eFEDS candidate location, but it is a defined as point source by XCS in the higher signal to noise {\em XMM data}. There are no associated galaxies in the SDSS or HSC images. \\ 
\hline
8922$^\dagger$ & 134.067 & -1.663 & 0.514 & Spurious: In eFEDS, this is a spurious detection of the outskirts of the emission from an X-ray bright spiral galaxy. In the higher resolution {\em XMM} image, there is no source at this location.\\ 
\hline
9463 & 136.753 & 1.176 & 0.799 & Blend: In the higher signal to noise (18501 s exposure) {\em XMM} image, two point sources are detected. \\ 
\hline
13484 & 136.766 & 1.132 & 0.307 & Spurious: There is no obvious extended X-ray emission in either the {\em eROSITA} or {\em XMM} data.  There are no associated galaxies in the SDSS or HSC images. \\ 
\hline
13299 & 138.691 & 4.439 & 0.348 & Spurious: In eFEDS, this is a spurious detection of the outskirts of the emission from an X-ray bright star. \\ 
\hline
11754 & 140.018 & 1.007 & 0.033 & Spurious: This is a spurious detection in the outskirts of the nearby eFEDS candidate ID 150 ($z$=0.017). \\ 
\hline
5702 & 130.295 & 0.867 & 0.415 & Spurious: In the higher signal to noise (130445 s exposure) {\em XMM} image, a point source is detected which appears to be associated with a  blue (i.e. likely AGN) object in the SDSS and HSC images.\\
\hline
6840 & 135.597 & 1.868 & 0.561 & Spurious: In the {\em XMM} image, XCS detects a point source which appears to be associated with a  blue (i.e. likely AGN) object in the SDSS and HSC images. \\ 
\hline
\end{tabular}
\end{center}
\end{table*}

\begin{table*}
\begin{center}
\caption[]{{eFEDS-{\em XMM} galaxy cluster candidates which appear to be galaxy clusters whose X-ray emission is significantly contaminated by another source. \newline $^\dagger$ indicates that the candidate was present in the optically confirmed sample from \cite{efedsclusteropticalcat}.}\label{tab:taintedxray}}
\vspace{-5mm}
\begin{tabular}{cccc|p{0.70\linewidth}}
\hline
\hline
eFEDS ID & RA & Dec & $z$ & Notes\\
\hline
\hline
16370$^\dagger$ & 134.098 & -1.604 & 0.425 & eFEDS candidate coincident with a collection of galaxies in SDSS/HSC; however X-ray emission is contaminated by low redshift spiral galaxy. \\ 
\hline
150$^\dagger$ & 140.009 & 1.039 & 0.017 & SDSS/HSC indicates the presence of a group of galaxies, however, the X-ray emission originates primarily from the central galaxy. \\ 
\hline
3133$^\dagger$ & 140.649 & -0.412 & 0.055 & SDSS/HSC indicates the presence of a group of galaxies, however, the X-ray emission originates primarily from the central galaxy. \\ 
\hline
3008$^\dagger$ & 130.451 & 0.82 & 0.078 & SDSS/HSC indicates the presence of a group of galaxies, however, the X-ray emission originates primarily from the central galaxy. \\ 
\hline
\end{tabular}
\end{center}
\end{table*}

\section{\lowercase{e}FEDS Candidate 1023}
\label{app:split_cluster}

This galaxy cluster has been been split into two sources by the eFEDS source finder; a visual inspection confirmed a single extended X-ray source (in both {\em eROSITA} and {\em XMM} images) and a single projected distribution of red galaxies (Section~\ref{subsubsec:frag}). XCS also detected this as a single extended source.

One of the two eFEDS catalogue entries that make up this cluster has measured {\em eROSITA} $T_{\rm{X}}$ and $L_{\rm{X}}$ values. These values will be impacted by the masking of emission from the other component. Therefore it would not be appropriate to include those values in the comparisons presented in Figures~\ref{fig:l500kpccomp} or ~\ref{fig:t500kpccomp}.

However, we have attempted to mimic the {\em eROSITA} values using {\em XMM} data. This involves manually adding a region to be excluded when the {\em XMM} spectra are generated. This region is centered on the eFEDS X-ray candidate catalogue coordinates for eFEDS-8602, and uses the `extent' value for that candidate published in the optical counterpart catalogue as the radius of the new exclusion region. Doing this, we find the {\em XMM} determined $L_{\rm{X}}$ and $T_{\rm{X}}$ values are consistent with those presented in \cite{efedsclustercat}; see Figures~\ref{fig:l500kpccomp} and ~\ref{fig:t500kpccomp} (cyan diamond). 

\section{\lowercase{e}FEDS-XCS Data and Measurements}
\label{app:measurements}
In Table \ref{tab:xmmobs} we present information on the {\em XMM} data that were used for each eFEDS-XCS cluster, including the unique {\em XMM} observation identifier and which instruments had usable data. We also include information on which instruments of which observations were contributed to the final luminosity and temperature measurements of each eFEDS-XCS cluster. 
In Table \ref{tab:measurements} we present temperature and luminosity values measured for the clusters in the eFEDS-XCS sample. We use \texttt{XGA} to generate spectra and run XSPEC fits for these clusters. The fitting procedure is discussed in more detail in Section \ref{subsec:fitproc}. All measurements are centered on the eFEDS coordinates from the X-ray cluster candidate catalogue, with redshift information also taken from that catalogue.

\begin{table*}
\begin{center}
\caption{{The {\em XMM} data used in the analysis of the eFEDS-XCS sample, individual clusters denoted by their unique eFEDS ID. ObsID contains the unique identifier(s) of the {\em XMM} observation(s) used. T denotes true, F denotes false, - denotes that either no successful spectral fit was performed, or the data for that cameras was not available. Columns with a subscript A (e.g. $\rm{PN}_{\rm{A}}$) indicate whether that instrument is available for an ObsID. Columns with a subscript radius (e.g. $\rm{PN}_{\rm{500kpc}}$) indicate whether that instrument's data contributed to the final XSPEC fit from which we extract temperature and luminosity information.}\label{tab:xmmobs}}
\vspace{1mm}
\begin{tabular}{ccccccccccc}
\hline
\hline
eFEDS ID & ObsID & PN$_{\rm{A}}$ & MOS1$_{\rm{A}}$ & MOS2$_{\rm{A}}$ & PN$_{\rm{500kpc}}$ & MOS1$_{\rm{500kpc}}$ & MOS2$_{\rm{500kpc}}$\\
\hline
\hline
\multirow{2}{4em}{\centering 6605} & 0202940101 & T & T & T & F & F & F \\ & 0202940201 & T & T & T & T & T & T \\ 
\hline
144 & 0650381601 & T & T & T & T & T & T \\ 
\hline
7831 & 0784350101 & T & T & T & T & T & T \\ 
\hline
1023 & 0761730501 & T & F & T & T & - & T \\ 
\hline
6125 & 0761730501 & T & F & T & T & - & T \\ 
\hline
339 & 0655340137 & T & T & T & T & T & T \\ 
\hline
4810 & 0655340137 & T & T & T & T & F & F \\ 
\hline
1458 & 0655340135 & T & T & T & T & T & T \\ 
\hline
2079 & 0651170301 & T & T & T & T & T & T \\ 
\hline
569 & 0783881001 & T & T & T & T & T & T \\ 
\hline
1385 & 0783881001 & T & F & T & T & - & T \\ 
\hline
8857 & 0725290142 & T & T & T & - & - & - \\ 
\hline
\multirow{3}{4em}{\centering 3171} & 0725300134 & T & F & T & F & - & T \\ & 0725290144 & T & T & T & T & T & F \\ & 0725290145 & T & F & T & T & - & T \\ 
\hline
\multirow{2}{4em}{\centering 8881} & 0725290139 & T & T & T & F & F & F \\ & 0725290146 & T & T & T & T & F & T \\ 
\hline
\multirow{2}{4em}{\centering 1104} & 0655340160 & T & T & T & T & F & T \\ & 0804410201 & T & T & T & T & T & T \\ 
\hline
\multirow{2}{4em}{\centering 4232} & 0655340160 & T & T & T & F & F & F \\ & 0804410201 & T & T & T & T & T & T \\ 
\hline
\multirow{3}{4em}{\centering 5655} & 0725300158 & T & T & T & T & T & F \\ & 0725300159 & T & F & T & T & - & T \\ & 0725300136 & T & T & F & T & F & - \\ 
\hline
\multirow{3}{4em}{\centering 1712} & 0725300140 & T & T & F & T & T & - \\ & 0725300132 & T & T & T & T & F & F \\ & 0725300131 & T & T & T & F & F & T \\ 
\hline
\multirow{2}{4em}{\centering 5774} & 0725300157 & T & T & T & - & - & - \\ & 0725310131 & T & T & T & - & - & - \\ 
\hline
\multirow{2}{4em}{\centering 3590} & 0725310152 & T & T & T & T & F & F \\ & 0725300160 & T & F & T & F & - & F \\ 
\hline
\multirow{2}{4em}{\centering 12660} & 0725290131 & T & T & T & T & T & T \\ & 0725290154 & T & T & T & F & T & T \\ 
\hline
\multirow{3}{4em}{\centering 3585} & 0725310149 & T & T & T & T & F & F \\ & 0725310150 & T & T & T & T & F & T \\ & 0725310131 & T & F & F & T & - & - \\ 
\hline
\multirow{2}{4em}{\centering 5170} & 0725300152 & T & F & T & T & - & F \\ & 0725300153 & T & T & T & T & F & F \\ 
\hline
\multirow{2}{4em}{\centering 9359} & 0725300145 & T & F & T & T & - & T \\ & 0725300146 & T & T & T & T & F & F \\ 
\hline
\end{tabular}
\end{center}
\end{table*}

\begin{table*}
\begin{center}
\contcaption{{}}
\vspace{1mm}
\begin{tabular}{ccccccccccc}
\hline
\hline
eFEDS ID & ObsID & PN$_{\rm{A}}$ & MOS1$_{\rm{A}}$ & MOS2$_{\rm{A}}$ & PN$_{\rm{500kpc}}$ & MOS1$_{\rm{500kpc}}$ & MOS2$_{\rm{500kpc}}$\\
\hline
\hline
\multirow{2}{4em}{\centering 3259} & 0725300144 & T & T & T & T & T & T \\ & 0725300151 & T & T & T & F & F & T \\ 
\hline
\multirow{5}{4em}{\centering 7086} & 0725310133 & T & F & F & - & - & - \\ & 0725310147 & T & F & T & - & - & - \\ & 0725310148 & T & T & T & - & - & - \\ & 0725310157 & F & F & T & - & - & - \\ & 0402780801 & T & T & T & - & - & - \\ 
\hline
\multirow{3}{4em}{\centering 5219} & 0725310158 & T & T & T & T & F & F \\ & 0725310147 & T & F & F & T & - & - \\ & 0725310159 & T & T & T & T & F & F \\ 
\hline
7084 & 0725310157 & T & T & T & - & - & - \\ 
\hline
\multirow{2}{4em}{\centering 885} & 0725310142 & T & T & T & T & T & T \\ & 0725310141 & T & F & T & T & - & T \\ 
\hline
2004 & 0800400501 & T & F & T & T & - & T \\ 
\hline
3523 & 0800400501 & T & T & T & F & T & T \\ 
\hline
372 & 0602830401 & T & F & F & T & - & - \\ 
\hline
4253 & 0602830401 & T & F & T & F & - & T \\ 
\hline
\multirow{2}{4em}{\centering 534} & 0804410101 & T & T & T & T & T & T \\ & 0650381801 & T & T & T & T & T & T \\ 
\hline
100 & 0804410501 & T & T & T & T & T & T \\ 
\hline
857 & 0823710301 & T & T & T & T & T & T \\ 
\hline
12565 & 0802220601 & T & F & F & - & - & - \\ 
\hline
\end{tabular}
\end{center}
\end{table*}

\begin{table*}
\begin{center}
\caption[]{{eFEDS-XCS galaxy cluster \texttt{XGA} measured values, RA, Dec, and redshift are taken from the eFEDS X-ray cluster candidate catalogue. $T^{\rm{XGA}}_{\rm{x}, 500kpc}$ are temperatures within 500\,kpc apertures, given in keV. $L^{\rm{XGA, 52}}_{\rm{x, 500kpc}}$ and $L^{\rm{XGA, bol}}_{\rm{x, 500kpc}}$ are 0.5-2 .0keV and bolometric luminosities within a 500\,kpc apertures, in units of 10$^{43}$erg $\rm{s}^{-1}$. All uncertainties calculated from 68\% confidence limits, equivalent to $1\sigma$.}\label{tab:measurements}}
\vspace{1mm}
\begin{tabular}{cccccccccc}
\hline
\hline
eFEDS ID & RA & Dec & z & T$^{\rm{XGA}}_{\rm{x}, 500kpc}$ & L$^{\rm{XGA, 52}}_{\rm{x, 500kpc}}$ & L$^{\rm{XGA, bol}}_{\rm{x, 500kpc}}$\\
\hline
\hline
6605 & 130.353 & 0.777 & 0.41 & $1.61^{+0.17}_{-0.13}$ & $2.32^{+0.16}_{-0.17}$ & $4.81^{+0.46}_{-0.41}$ \\ 
\hline
144 & 131.37 & 3.461 & 0.33 & $5.8^{+0.59}_{-0.48}$ & $19.18^{+0.84}_{-0.67}$ & $67.1^{+3.88}_{-3.88}$ \\ 
\hline
7831 & 132.272 & 2.243 & 0.4 & $1.61^{+0.1}_{-0.11}$ & $2.82^{+0.21}_{-0.15}$ & $5.87^{+0.44}_{-0.43}$ \\ 
\hline
1023$^{\dagger}$ & 132.616 & 0.251 & 0.2 & $3.08^{+0.14}_{-0.13}$ & $3.55^{+0.04}_{-0.05}$ & $9.29^{+0.25}_{-0.22}$ \\ 
\hline
6125 & 132.627 & 0.558 & 0.19 & $2.39^{+0.33}_{-0.31}$ & $0.98^{+0.05}_{-0.05}$ & $2.35^{+0.2}_{-0.18}$ \\ 
\hline
339 & 133.071 & -1.025 & 0.46 & $5.57^{+0.81}_{-0.63}$ & $18.3^{+0.71}_{-0.76}$ & $62.82^{+4.09}_{-6.48}$ \\ 
\hline
4810 & 133.13 & -1.208 & 0.55 & $2.06^{+2.12}_{-0.72}$ & $4.56^{+1.06}_{-1.85}$ & $10.14^{+3.84}_{-4.1}$ \\ 
\hline
1458 & 133.554 & -2.357 & 0.38 & $4.47^{+0.78}_{-0.72}$ & $11.59^{+0.53}_{-0.75}$ & $35.63^{+3.34}_{-4.39}$ \\ 
\hline
2079 & 133.696 & -1.359 & 0.35 & $3.27^{+0.56}_{-0.38}$ & $4.59^{+0.23}_{-0.27}$ & $12.23^{+0.86}_{-0.98}$ \\ 
\hline
569 & 134.086 & 1.78 & 0.72 & $4.94^{+0.16}_{-0.16}$ & $29.96^{+0.55}_{-0.44}$ & $96.29^{+2.02}_{-1.94}$ \\ 
\hline
1385 & 134.113 & 1.705 & 0.73 & $6.47^{+1.49}_{-1.11}$ & $7.06^{+0.4}_{-0.37}$ & $26.2^{+2.66}_{-3.5}$ \\ 
\hline
8857 & 134.658 & 1.449 & 0.75 & - & - & - \\ 
\hline
3171 & 135.269 & 1.279 & 0.25 & $1.7^{+0.24}_{-0.2}$ & $2.64^{+0.26}_{-0.63}$ & $5.69^{+0.76}_{-1.3}$ \\ 
\hline
8881 & 135.314 & 0.844 & 0.31 & $3.27^{+4.56}_{-1.79}$ & $1.13^{+0.34}_{-0.36}$ & $3.02^{+1.2}_{-1.7}$ \\ 
\hline
1104 & 135.372 & -1.648 & 0.31 & $5.96^{+0.32}_{-0.32}$ & $10.94^{+1.11}_{-0.81}$ & $38.75^{+4.67}_{-4.06}$ \\ 
\hline
4232 & 135.443 & -1.632 & 0.29 & $1.95^{+0.53}_{-0.3}$ & $0.99^{+0.1}_{-0.14}$ & $2.19^{+0.36}_{-0.33}$ \\ 
\hline
5655 & 135.735 & 1.774 & 0.12 & $0.97^{+0.11}_{-0.12}$ & $0.02^{+0.11}_{-0.02}$ & $0.04^{+0.19}_{-0.04}$ \\ 
\hline
1712 & 135.74 & 0.805 & 0.52 & $2.08^{+1.74}_{-0.63}$ & $6.12^{+2.32}_{-1.85}$ & $13.68^{+4.61}_{-4.53}$ \\ 
\hline
5774 & 136.036 & 1.432 & 0.84 & - & - & - \\ 
\hline
3590 & 136.078 & 2.112 & 0.81 & $4.99^{+8.64}_{-2.3}$ & $10.17^{+1.35}_{-3.57}$ & $32.84^{+7.6}_{-13.35}$ \\ 
\hline
12660 & 136.08 & -1.077 & 0.31 & $2.68^{+1.51}_{-0.81}$ & $0.81^{+0.21}_{-0.14}$ & $2.01^{+0.55}_{-0.47}$ \\ 
\hline
3585 & 136.42 & 1.539 & 0.64 & $4.0^{+1.46}_{-0.94}$ & $8.81^{+2.0}_{-1.89}$ & $25.52^{+6.27}_{-6.63}$ \\ 
\hline
5170 & 136.473 & 0.379 & 0.37 & $1.22^{+0.14}_{-0.18}$ & $2.33^{+0.38}_{-0.46}$ & $4.42^{+0.74}_{-0.94}$ \\ 
\hline
9359 & 136.502 & -0.423 & 0.3 & $1.92^{+1.46}_{-0.51}$ & $1.0^{+0.66}_{-0.67}$ & $2.2^{+1.89}_{-1.4}$ \\ 
\hline
3259 & 136.504 & 0.015 & 0.2 & $2.53^{+1.06}_{-0.59}$ & $1.22^{+0.33}_{-0.19}$ & $2.96^{+0.87}_{-0.68}$ \\ 
\hline
7086 & 136.654 & 1.148 & 0.79 & - & - & - \\ 
\hline
5219 & 136.977 & 0.961 & 0.74 & $4.74^{+2.17}_{-1.44}$ & $9.92^{+1.76}_{-1.6}$ & $31.21^{+5.25}_{-7.57}$ \\ 
\hline
7084 & 137.025 & 1.331 & 0.66 & - & - & - \\ 
\hline
885 & 137.308 & -0.204 & 0.31 & $2.99^{+0.52}_{-0.37}$ & $5.65^{+0.49}_{-0.44}$ & $14.52^{+1.38}_{-1.55}$ \\ 
\hline
2004 & 137.314 & -1.018 & 0.82 & $5.57^{+4.09}_{-1.96}$ & $6.17^{+2.72}_{-3.36}$ & $21.12^{+10.09}_{-10.65}$ \\ 
\hline
3523 & 137.386 & -0.839 & 1.13 & $5.61^{+2.19}_{-1.11}$ & $26.33^{+2.91}_{-3.09}$ & $93.83^{+11.49}_{-17.28}$ \\ 
\hline
\end{tabular}
\end{center}
\end{table*}

\begin{table*}
\begin{center}
\contcaption{{}}
\vspace{1mm}
\begin{tabular}{cccccccccc}
\hline
\hline
eFEDS ID & RA & Dec & z & T$^{\rm{XGA}}_{\rm{x}, 500kpc}$ & L$^{\rm{XGA, 52}}_{\rm{x, 500kpc}}$ & L$^{\rm{XGA, bol}}_{\rm{x, 500kpc}}$\\
\hline
\hline
372 & 138.723 & 4.27 & 0.14 & $2.46^{+0.84}_{-0.67}$ & $1.8^{+0.11}_{-0.09}$ & $4.35^{+0.5}_{-0.4}$ \\ 
\hline
4253 & 138.801 & 4.585 & 0.36 & $5.18^{+-5.18}_{-3.76}$ & $1.18^{+0.3}_{-1.18}$ & $3.96^{+5.96}_{-1.84}$ \\ 
\hline
534 & 139.042 & -0.397 & 0.32 & $6.04^{+0.35}_{-0.35}$ & $13.86^{+0.32}_{-0.34}$ & $49.42^{+1.96}_{-1.85}$ \\ 
\hline
100 & 140.338 & 3.291 & 0.33 & $7.83^{+0.31}_{-0.31}$ & $19.66^{+0.25}_{-0.23}$ & $79.71^{+0.85}_{-1.93}$ \\ 
\hline
857 & 140.55 & -0.459 & 0.32 & $5.86^{+0.24}_{-0.23}$ & $9.51^{+0.13}_{-0.11}$ & $33.41^{+0.66}_{-0.84}$ \\ 
\hline
12565 & 142.473 & 0.467 & 0.15 & - & - & - \\ 
\hline
\end{tabular}
\end{center}
\end{table*}



\bsp	
\label{lastpage}
\end{document}